\documentclass[10pt,twocolumn]{IEEEtran}
\usepackage{xcolor}

\usepackage{multicol}

\usepackage{color}
\usepackage{xcolor}
\usepackage{soul}
\usepackage{amsmath}
\usepackage{graphics}
\usepackage{setspace}
\usepackage{cite}
\usepackage{latexsym}
\usepackage{float}
\usepackage{epsfig}
\usepackage{multirow}
\usepackage{cite,cases,url}
\usepackage{amssymb}
\usepackage{graphicx}
\usepackage{epstopdf}
\usepackage{balance}
\usepackage{subcaption}
\usepackage{enumerate}
\usepackage{array}
\usepackage{enumitem}
\usepackage{dblfloatfix}
\usepackage[normalem]{ulem}
\useunder{\uline}{\ul}{}
\usepackage{algorithm,amsmath}
\usepackage{enumitem}
\usepackage{caption}
\usepackage{subcaption}
\usepackage{tikz}
\usepackage{adjustbox}
\makeatletter
\newcommand{\ssymbol}[1]{^{\@fnsymbol{#1}}}
\usepackage{etoolbox}
\usepackage[noend]{algpseudocode}
\usepackage{tikzpagenodes}      
\usepackage[normalem]{ulem}
\useunder{\uline}{\ul}{}
\usepackage[english]{babel}
\usepackage{amsthm}
\usepackage{tabularx}
\newcolumntype{L}{>{\centering\arraybackslash}m{5cm}}
\newcolumntype{K}{>{\centering\arraybackslash}m{6cm}}
\newcolumntype{P}{>{\centering\arraybackslash}m{2.3cm}}
\newcolumntype{M}{>{\raggedright\arraybackslash}m{2cm}}
\newcolumntype{N}{>{\raggedright\arraybackslash}m{2.5cm}}

\usepackage{changepage}
\usepackage{xpatch}
\xpatchcmd\algorithmic
  {\labelwidth 1.2em}
  {\labelwidth .7em}
  {}{\fail}
\usepackage{physics}
\makeatletter
\def\thanks#1{\protected@xdef\@thanks{\@thanks
        \protect\footnotetext{#1}}}
\makeatother

\newtheorem{theorem}{Theorem}[section]

\newtheorem{lemma}[theorem]{Lemma}
\makeatletter
\newcommand{\multiline}[1]{%
  \begin{tabularx}{\dimexpr\linewidth-\ALG@thistlm}[t]{@{}X@{}}
    #1
  \end{tabularx}
}
\makeatother
\begin{document}
\title{
UAV Trajectory and Multi-User Beamforming Optimization for Clustered Users Against Passive Eavesdropping Attacks With Unknown CSI}
\author{\IEEEauthorblockN {Aly Sabri Abdalla, Ali Behfarnia, and Vuk Marojevic} 
\thanks{Manuscript received 8 September 2022; revised 3 November
2022; accepted 4 June 2023. Date of publication - - 2023; date of
current version 10 June 2023. This work was supported in part by the NSF PAWR program, under grant number CNS-1939334. The work of A. Behfarnia is supported in part by the Faculty Research Program under the Office of Sponsored Programs and Research at the University of Tennessee at Martin.\\
Aly Sabri Abdalla and Vuk Marojevic are with the Department of Electrical and Computer Engineering, Mississippi State University, MS, USA e-mail: ($asa298$@msstate.edu; $vm602$@msstate.edu).\\
Ali Behfarnia is with the Department of Engineering, University of Tennessee at Martin, TN, USA e-mail: (a.behfarnia@tennessee.edu).
}
}
\date{}
\vspace{-2 cm}
\maketitle
\begin{tikzpicture}[remember picture,overlay]
   
    \node[align=center,text=black] at ([yshift=-2em]current page text area.south) {\footnotesize Copyright (c) 2015 IEEE. Personal use of this material is permitted. However, permission to use this material for\\[-0.3 em]
    \footnotesize any other purposes must be obtained from the IEEE by sending a request to pubs-permissions@ieee.org.};
  \end{tikzpicture}%

\begin{abstract}
This paper tackles the fundamental passive eavesdropping problem in modern wireless communications in which the location and the channel state information (CSI) of the attackers are unknown. In this regard, we propose deploying an unmanned aerial vehicle (UAV) that serves as a mobile aerial relay (AR) to help ground base station (GBS) support a subset of vulnerable users. More precisely, our solution (1) clusters the single-antenna users in two groups to be either served by the GBS directly or via the AR, (2) employs optimal multi-user beamforming to the directly served users, and (3) optimizes the AR's 3D position, its multi-user beamforming matrix and transmit powers by combining closed-form solutions with machine learning techniques. Specifically, we design a plain beamforming and power optimization combined with a deep reinforcement learning (DRL) algorithm for an AR to optimize its trajectory for the security maximization of the served users. Numerical results show that the multi-user multiple input, single output (MU-MISO) system split between a GBS and an AR with optimized transmission parameters without knowledge of the eavesdropping channels achieves high secrecy capacities that scale well with increasing the number of users.

\end{abstract}

\IEEEpeerreviewmaketitle

\section{Introduction}
\label{sec:intro}

 \color{black} Unmanned aerial vehicles (UAVs) are envisioned to improve the next generation of wireless communication systems, 6G and beyond, by providing flexible, intelligent, secure, and limitless connectivity \cite{mozaffari21, tataria20216g,Sec}. Steps to identify the challenges and solutions of emerging cellular networks to serve UAVs are being undertaken by the 3rd Generation Partnership Project (3GPP)~\cite{3GPP_Stnds}. 
 A prominent use case for a UAV is the aerial relay (AR) which 
supports extended coverage or higher system capacity at low-cost. 
However, the largely line of sight (LoS) air-to-ground (A2G) communications between UAVs and user equipment (UEs) 
make the system vulnerable to a variety of attacks \cite{fotouhi2019survey}. Eavesdropping is a major passive attack that can compromise communications channels and gain access to private and sensitive user information.

Physical layer security has been introduced as a powerful tool to secure communication links by using the physical characteristics  of wireless communication channels \cite{VincentPoor2017PNAS,  PLS16_2, PLS22_1}. UAVs can benefit from physical layer security by applying the latest technologies such as artificial intelligent (AI) methods as well as various communication techniques to mitigate the compromise of malicious  behavior in the network. However, applying such techniques is complicated by three factors: (i) requiring to coordinate between a ground base station (GBS) and the AR to determine which users should be served by which stations, (ii) handling resource limitations and trajectory of ARs to serve specific UE(s), and (iii) choosing the well-suited learning and communication techniques for the GBS and the AR to dynamically maximize the security metrics.

 \color{black}

\color{black}

\subsection{Related Work}

The recent related works 
can be classified into three
groups
: i) beamforming-aided secure communications; ii) UAV-aided secure communications; iii) beamforming and UAV-aided secure communications. 
Table~\ref{tab:priorart} provides a summery of the prior art and proposed research related to the work presented in this paper.
\setlist[itemize]{leftmargin=3.5mm}

    \begin{table*}[ht]
\centering
\caption{\textcolor{black}{Prior Art and Proposed Research.}}
\small
{\begin{tabular}{|p{1.7cm}|p{0.6cm}|p{2.7cm}|p{3cm}|p{4.9cm}|p{2.1cm}|}
\hline
 \textbf{Category} &\textbf{Ref.} &\textbf{Objective Metric} 
 &\textbf{Attack type}  &\textbf{Strategy} &\textbf{
 Attackers' CSI} 
\\ \hline

Beamforming
& \cite{BaiBF}
& Secrecy sum rate 
& Passive eavesdroppers
& RCI is adopted to drive the power allocation between AN and information signal. 
& Perfect CSI with channel errors\\
\cline{2-6}
& \cite{MIMOBF}
& Secrecy rate  
& Active eavesdropper
& Analytical framework to find the best combination of AN and beamforming. 
& Perfect CSI\\
\cline{2-6}
& \cite{MISOBF}
& Secrecy rate   
& Active and passive eavesdroppers
& Analytical framework to design the beamforming. 
& Statistical CSI
\\ \hline \hline

UAV 
& \cite{UAVSEC1}
& Average secrecy rate 
& Passive eavesdropper
& Optimizing the UAV's trajectory and AN allocation via iterative algorithm.
& Perfect CSI\\
\cline{2-6}
& \cite{UAVSEC2}
& Secrecy rate   
& Passive eavesdropper
& Jointly optimizing the source/ UAV relay transmit power
and the UAV trajectory through an iterative algorithm.
& Perfect CSI\\
\cline{2-6}
& \cite{AlyWCNC}
& Secrecy rate  
& Passive eavesdropper
& The UAV’s trajectory and transmit power allocation are jointly optimized by applying DQL algorithm.
& 
Unknown CSI
\\ \hline \hline

Beamforming and UAV
& \cite{UAVBF1}
& Minimum secrecy rate
& Passive eavesdropper
& Jointly optimizing the UAV beamforming and position to enhance UE's secrecy rate through applying multi-objective dragonfly algorithm.
& Perfect CSI\\
\cline{2-6}
& \cite{UAVBF2}
& Secrecy capacity  
& Passive eavesdroppers
&  A DRL is proposed to optimize the UAV trajectory and transmitter and jammer UAVs beamforming. 
& Perfect CSI\\
\cline{2-6}
& \textcolor{black}{\cite{UAVSatSec}}
& \textcolor{black}{Secrecy rate } 
& \textcolor{black}{Passive eavesdropper}
&  \textcolor{black}{Jointly optimize the beamforming of multi-beam satellite
and the power allocation of UAV through an iterative alternating optimization approach.}
& \textcolor{black}{Perfect CSI}
\\ \hline \hline

This work
& 
& Secrecy sum capacity 
& Passive eavesdroppers
& User clustering for 
association with the GBS and AR, where a DQL is designed to optimize the UAV trajectory, beamforming, and power control without knowledge of the wiretap CSI. 
& 
Unknown CSI
\\ \hline 

\end{tabular}%
}

\label{tab:priorart}
\end{table*}

    \textbf{Beamforming-aided Secure Communications:} 
    Transmit beamforming limit the radio frequency (RF) propagation footprint and thus implicitly enable a secure the propagation channel without high computational requirements a the receiver as compared to cryptographic security schemes. Carefully designing the beam patterns of antennas at the transmitter, receiver, or both can enhancing system performance and security parameter, such as signal-to-interference plus noise ratio (SINR) and secrecy rate, respectively. Researchers have studied how to leverage and optimize beamforming for improving the PLS of current and future wireless communication networks. The work presented in~\cite{BaiBF} investigates the achieved secrecy sum rate for a multi-cell multiple-input multiple-output (MIMO) system which is under a passive eavesdropper attack. The power allocation between artificial noise (AN) and information signal is managed to maximize the sum secrecy rate with imperfect channel state information (CSI), which is derived using regularized channel inversion (RCI) precoding. Reference~\cite{MIMOBF} proposes strategies of combining AN and beamforming to achieve high secrecy performance for massive MIMO systems in spite of single-antenna active eavesdropping attacks that attempt to spoil the channel estimation acquisition at the BS. 
   Reference~\cite{MISOBF} derives the  multiple-input, single-output (MISO) beamforming design for random wireless networks with statistical CSI in an environment with eavesdroppers and interferers. 
    
    \textbf{UAV-aided Secure Communications:} 
    Reference~\cite{UAVSEC1} demonstrates the applicability of maximizing the achievable average secrecy rate by optimizing the AN transmission 
    and UAV trajectory. 
    Reference~\cite{UAVSEC2}  proposes using the UAV as a relay 
    to improve the secrecy rate by jointly optimizing the source/relay transmit power and the UAV trajectory. In our previous work~\cite{AlyWCNC}, we have proposed a deep Q-learning (DQL) algorithm to optimize the secrecy rate by optimizing the trajectory of the UAV relay and the transmit power without the availability of the CSI of the wiretap channel. 
    
    \textbf{Beamforming Plus UAV-Aided Secure Communications:} Combining both beamforming and UAV has been considered as an enhanced PLS technique in advanced wireless communications. For example, \cite{UAVBF1} introduces the multi-objective dragonfly algorithm (MODA) to solve the multi-objective optimization problem for enhancing the minimum secrecy rate between the UAV node and a single UE for different clusters. The work presented in that paper assumes perfect 
    CSI conditions
    for the BS and focuses on optimizing the UAV performance. Reference~\cite{UAVBF2}  proposes a multi-agent deep reinforcement learning (DRL) algorithm to maximize the secrecy capacity of a multi user system 
    by optimizing the trajectory of the aerial BS  and the beamforming matrix of the jammer UAV 
    interfering with the eavesdroppers. 
    \textcolor{black}{The authors of~\cite{UAVSatSec} propose an iterative optimization approach that alternately optimizes the beamforming of satellite transmitters and the power allocation of the UAV acting as an aerial relay and friendly jammer supporting 
 multi-beam satellite-enabled vehicle communication in the presence of eavesdropping.}

\subsection{Contribution}
In this paper, we aim to 
\color{black}
mitigate passive eavesdropping attacks, where an eavesdropper illegitimately wiretaps the legitimate wireless communications links. 
 \color{black}
To this end, we propose a combination of machine learning, deep reinforcement learning, and multi-antennas techniques at the BS and the AR to maximize the security of UEs in a wireless communication network. 
\color{black} 
The contributions of this paper are:
\begin{itemize}
    \item We define a practical optimization problem to maximize the channel secrecy capacity without CSI knowledge of the wiretap channel.
    \item We introduce a framework for effectively solving this problem by means of user clustering, beamforming and power control, and AR trajectory optimization. We design a DRL solution for the trajectory optimization and leverage the closed form solutions for the beamforming and transmit and relay power allocation. 
    \item We provide a comprehensive numerical analysis that demonstrates the effectiveness of the proposed tools. 
\end{itemize}

The rest of paper is organized as follows. Section II presents the system model. 
Section III formulates the problem and defines the relevant metrics. 
Section IV derives the 
solution. Numerical results and analyses are presented in Section V. Section VI provides the concluding remarks. 
\color{black}

\label{sec:related}

\section{\color{black}System Model} 
\vspace{-1 mm}
\label{sec:system}
\color{black}
We consider 
a ground base station (GBS) 
serving ground UEs where the communication links are subject to passive eavesdropping attacks. 
The eavesdroppers have a radio receiver and can wiretap the downlink transmission. A UAV acting as an 
AR is dispatched to support secure communications. This scenario is illustrated in Fig. 1. 
\begin{figure}[ht]
	   \centering
	   \includegraphics[width=0.48\textwidth]{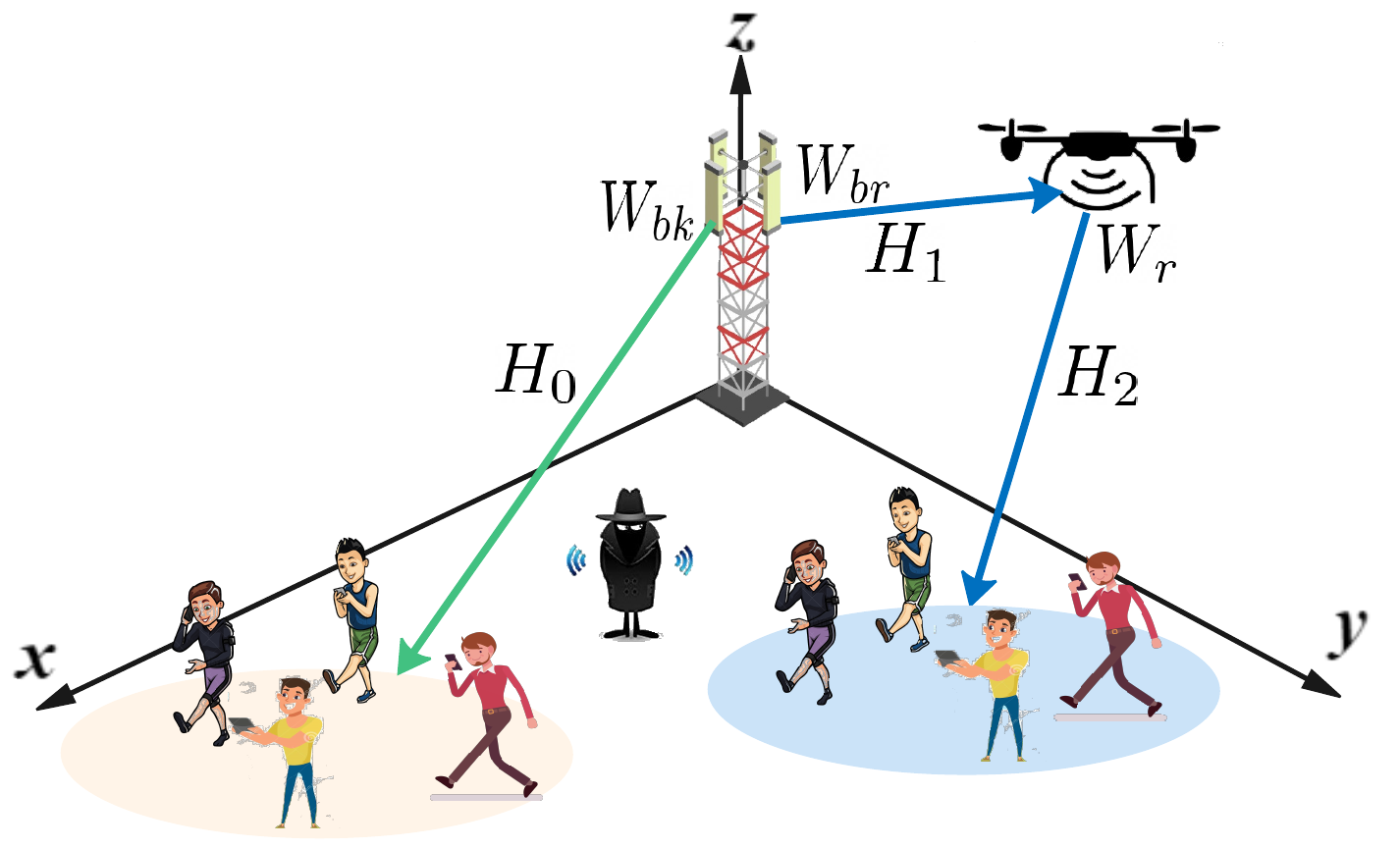}
        \caption{System model.
        }
        \vspace{-3 mm}
        \label{fig:simuSetup}
\end{figure}
\color{black}

We use the following notation: lower-case letters represent scalars and bold lower-case letters denote vectors. Bold upper-case letters are used for matrices. Tr($\mathbf{S}$) and $\mathbf{S}^{-1}$ represent the trace and the inverse of a square matrix $\mathbf{S}$, respectively. The operator (.)$^{T}$ denotes transpose, and the operator (.)$\ssymbol{2}$ denotes conjugate transpose. $\mathbf{S}(i,j)$ shows the $(i, j)$th element of matrix $\mathbf{S}$ and Rank($\mathbf{S}$) shows the rank of the matrix. $||\mathbf{v}||$ represents the Euclidean norm of a complex vector $\mathbf{v}$. Also, $|v|$ denotes the norm of a complex number $v$. $\mathbb{C}^{a \times b}$ denotes the dimension of $a \times b$ for a complex vector or matrix. Complex normal distribution vector with the mean vector $\mathbf{m}$ and the covariance matrix $\mathbf{\Sigma}$ is denoted by $ \mathcal{CN}(\textbf{m}, \mathbf{\Sigma})$, and $\sim$ implies "distributed as".

\subsection{Channel model}
\subsubsection{Air-to-ground}
\color{black}
In terms of modelling the A2G communication channel between the UAV and ground receivers, we consider small-scale Rician fading where the line of sight (LoS) component coexist with non-LoS (NLoS) components~\cite{A2GModel}. 
The GBS and AR have both a uniform
linear array (ULA) of $M$ and $N$ antennas, respectively. 
The A2G channel model, 
\vspace{-8 pt}
\begin{equation}\label{eq:A2G_channel}
\vspace{-4 pt}
\vb*{G}_{TR}=\frac{\sqrt{\lambda_0}}{d^{\alpha}_{TR}}\bigg( \sqrt{\frac{\beta}{1+\beta}} \ \vb*{G^{LoS}_{TR}} + \sqrt{\frac{1}{\beta+1}} \ \vb*{G^{NLoS}_{TR}} \bigg), 
\end{equation}
is obtained as the superposition of the LoS and NLoS channel components, where $\lambda_0$ is the path loss at the reference distance of $1 \ m$, $d_{TR}$ is the 3D distance between the GBS and AR
, $\alpha$ is the path loss exponent, and $\beta$ is the Rician factor. 
Without loss of generality, the entries of $\vb*{G^{NLoS}_{TR}}$ are assumed to be independent and identically distributed (i.i.d.) zero-mean and unit variance circularly symmetric complex Gaussian (CSCG), i.e., $\sim \mathcal{CN}(0,1)$. The LoS component, 
\vspace{-4 pt}
\begin{equation}
\vspace{-6 pt}
\vb*{G}^{LoS}_{TR}=\vb*{g^{(A)}_{TR}} \ \vb*{g^{(D)}_{TR}}, 
\end{equation}
where 
\vspace{-8 pt}
\begin{equation}
\vspace{-4 pt}
\vb*{g^{(A)}_{TR}}=\Big[ 1, e^{-j\frac{2\pi}{\lambda}\Upsilon\Lambda^{TR}}, \cdots, e^{-j\frac{2\pi}{\lambda}(N-1)\Upsilon\Lambda^{TR}} \Big]
\end{equation}
and
\vspace{-8 pt}
\begin{equation}
\vspace{-4 pt}
\vb*{g^{(D)}_{TR}}=\Big[ 1, e^{-j\frac{2\pi}{\lambda}\Upsilon\Gamma^{TR}}, \cdots, e^{-j\frac{2\pi}{\lambda}(M-1)\Upsilon\Gamma^{TR}} \Big] 
\end{equation}
correspond to channel contributions from the angel-of-arrival (AoA) and angel-of-departure (AoD) between the GBS and the AR. 
Parameter $\lambda$ is the carrier wavelength, $\Upsilon$ is the antenna separation, $\Lambda^{TR}= cos \ \Theta \ sin \ \varphi$ is the AoA component ($\Theta$--azimuth and $\varphi$--elevation AoA), and $\Gamma^{TR}= sin \ \vartheta \ cos \ \psi$ is the AoD component ($\vartheta$--elevation and $\psi$--azimuth AoD) of the transmitted signal from the GBS to the AR. 

The A2G channel between the 
AR and the 
ground users, 
\vspace{-6 pt}
\begin{equation}
\vspace{-4 pt}
\vb*{G}_{RK}=\frac{\sqrt{\lambda_0}}{\vb*{d^{\alpha}_{RK}}}\bigg( \sqrt{\frac{\beta}{1+\beta}} \ \vb*{g^{LoS}_{RK}} + \sqrt{\frac{1}{\beta+1}} \ \vb*{G^{NLoS}_{RK}} \bigg), 
\end{equation}
has an LoS and an NLoS term, where $\vb*{d_{RK}}$ is the 3D distance between the AR and the ground user cluster. 
The $\vb*{G^{NLoS}_{RK}}$ entries follow the same CSCG distribution 
as $\vb*{G^{NLoS}_{TR}}$. The LoS term, 
\vspace{-8 pt}
\begin{equation}
\vspace{-4 pt}
\vb*{g^{LoS}_{RK}}=\Big[ 1, e^{-j\frac{2\pi}{\lambda}\Upsilon\chi^{RK}}, \cdots, e^{-j\frac{2\pi}{\lambda}(N-1)\Upsilon\chi^{RK}} \Big], 
\end{equation}
defines the AoD components $\chi^{RK}= cos \ \Phi \ sin \ \Omega$  ($\Phi$--azimuth and $\Omega$--elevation AoD) of the transmitted signal from the ULA of the AR to the single-antenna users. 

\vspace{0.1cm}
\subsubsection{ Ground-to-ground}
\color{black}

the Alpha-beta-gamma (ABG)~\cite{ABG} channel model is adopted for the ground-to-ground (G2G) communication channels between the GBS and the eavesdropper and between the UEs and the eavesdropper. It is the closest path-loss model approximation to the actual 5G ground communications measurement results and it is employed by standard organizations such as ITU-R, 3GPP
, mmMAGIC, and QuaDRiGa~\cite{ABG2}. It is defined as 
\vspace{-5 pt}
\begin{equation}
\vspace{-4 pt}
\begin{aligned}
   h_{G2G}(f,d)&= 10 \hspace{3 pt} \rho_{G} \times 
log\Big(\frac{d_{gg}}{1 \hspace{1 pt} m}\Big)+\jmath\\
&+10\hspace{3 pt} \gamma_{G} \times 
log\Big(\frac{f_{c}}{1 \hspace{1 pt} GHz}\Big)+ \chi_{\sigma}^{G2G},
\end{aligned}
\end{equation}
where $d_{gg}$ is the 2D distance between the transmitter and receiver nodes, $\jmath$ is the intercept, and $\rho_{G}$ and $\gamma_{G}$ correspond to the distance and the frequency-dependent exponents. 
Shadow fading, 
$\chi_{\sigma}^{G2G}$, 
is modeled as a Gaussian random variable of zero-mean and standard deviation $\sigma_{sh}$. 


\color{black}
\subsection{Communication Model for Legitimate Users}
The $M$-antenna GBS can communicate with the $K$ single-antenna UEs either directly or using the $N$-antenna AR at the same frequency, employing space-division multiple access (SDMA) and time-division multiple access (TDMA) 
~\cite{zhang09}. 
\textcolor{black}{The GBS serves $K_b$ users directly 
and $K_r$ users via the AR, where $K=K_b+K_r$.} 
In what follows, we 
provide the corresponding communication models and channel capacities. 


\subsubsection{\color{black} Direct communication from GBS}
\color{black}
For the direct communication, the GBS 
forms multiple simultaneous beams to 
spatially separated users employing 
SDMA. 
The transmit beamforming assigns one beam vector for each user. However
, transmit power leakage can occur between beams causing multi-user interference. 

We consider the downlink transmission, where the GBS transfers $K_b$ data streams to $K_b$ users. The transmitted signal model is 
\vspace{-6 pt}
\begin{equation}
\vspace{-4 pt}
    \vb*{x}_{b}= \sum\limits_{k=1}^{K_b} \vb*{w}_{b,k} \, s_k, 
    \label{eq:xs}
\end{equation}
where $\vb*{x}_b \in \mathbb{C}^{M\times 1}$, $\vb*{w}_{b,k} \in \mathbb{C}^{M\times1}$ is the beamforming vector and $s_k$ the transmitted information symbol for the $k$th user. 
The beamforming, or precoding, matrix of the GBS 
contains $K_b$ beamforming vectors, 
$\vb*{W}_{bk} \in \mathbb{C}^{M \times K_b}$, where $\vb*{W}_{bk} =[\vb*{w}_{b,1}, \cdots, \vb*{w}_{b,K_b}]$.
\color{black}
The allocated transmit power for the $k$th user can then be calculated by the squared norm of the beamforming vector $\parallel\vb*{w}_{b,k} \parallel^2$.
\color{black} The received signal at the $K_b$ users can is expressed as 
\vspace{-6 pt}
\begin{align}\label{eq:direct_1}
\vspace{-6 pt}
\vb*{y}_0 = \vb*{H}_0 \, \vb*{x}_{b} \, + \, \vb*{n}_{0},
\end{align}
where $\vb*{y}_0 \in \mathbb{C}^{K_b\times 1}$, $\vb*{H}_0 \in \mathbb{C}^{K_b\times M}$ represents the channel between the $M$ antennas of the GBS 
and the $K_b$ single-antenna users, and $\vb*{n}_{0} \in \mathbb{C}^{K_b\times 1}$ represents noise. 
It is assumed that the distribution of noise at each user is complex normal with zero-mean and unit variance, i.e., 
$n_k \sim \mathcal{CN}(0, 1)$. The received signal at user $k$, 
\vspace{-6 pt}
\begin{align}\label{eq:direct_2}
\vspace{-6 pt}
\notag y_{0,k} &= \vb*{h}_{0,k} \, \vb*{x}_{b} \, + \, n_k,  \\[-4pt] 
\notag        &= \vb*{h}_{0,k} \Big( \sum\limits_{k=1}^{K_b} \vb*{w}_{b,k} \ s_k \Big) + \, n_k, \\[-10pt]
               &= \vb*{h}_{0,k} \vb*{w}_{b,k}\,s_k \, + \vb*{h}_{0,k} \Big( \sum\limits_{\substack{i=1 \\ i\neq k}}^{K_b}  \vb*{w}_{b,i} \, s_i \Big) \, + \, n_{k}\\[-21pt]
               \notag
\end{align}
has the signal-to-interference-plus-noise-ratio (SINR) 
\vspace{-8 pt}
\begin{align}\label{eq:direct_3}
    \gamma_{b,k}&=  \frac{\mid \vb*{h}_{0,k} \vb*{w}_{b,k} \mid^2}{\sum\limits_{{i\neq k}}\mid \vb*{h}_{0,k} \vb*{w}_{b,i} \mid^2 + 1} \,,\\[-21pt]
               \notag
\end{align}
where $\vb*{h}_{0,k} \in \mathbb{C}^{1\times M}$ denotes the MISO channel from the GBS to the $k$th user.
The channel capacity of 
the direct link 
is obtained from
\vspace{-8 pt}
\begin{align}\label{eq:direct_4}
C_{b,k} = \log_2 \big(1+\gamma_{b,k}\big).
\end{align}


\subsubsection{\color{black} Indirect communication via AR}\label{subsec:AR_model}

\textcolor{black}{We assume a time-slot based synchronization between the GBS transmission and the AR transmission~\cite{tdmaUAV, tdmaUAV2}. In odd time-slots (phase), the BS transmits $K_r$ data streams to the AR, each of which is destined to one UE.} \color{black} The transmission between the GBS and AR can be modeled as a standard point-to-to-point MIMO channel. The received signal at the AR \color{black} can be written as 
\vspace{-8 pt}
\begin{align}\label{eq:relay_1}
\notag   \vb*{y}_{1} &= \vb*{H}_1 \vb*{x}_b + \vb*{n}_{1},\\[-6pt]
                     &= \vb*{H}_1 \Big( \sum\limits_{k=1}^{K_r} \vb*{w}_{b,k} \ s_k \Big) + \, \vb*{n}_1,\\[-20pt]
               \notag
\end{align}
where $\vb*{y}_1\in \mathbb{C}^{N\times 1}$, $\vb*{H}_1 \in \mathbb{C}^{N\times M}$ is the MIMO communication channel between the BS and the AR, and $\vb*{n}_1 \sim \mathcal{CN}(0, \vb*{I}) \in \mathbb{C}^{N\times 1}$ is the noise vector. 

\color{black}
In the even time-slots, the AR transmits 
\vspace{-9 pt}
\begin{align}\label{eq:relay_2}
        \vb*{x}_r = \vb*{W}_r \ \vb*{y}_{1},
\end{align}
where $\vb*{x}_r \in \mathbf{C}^{N \times 1}$ and \color{black} 
$\vb*{W}_r \in \mathbb{C}^{N\times N}$ is the beamforming matrix.
The received signals at the $K_r$ UEs are modeled as 
\vspace{-9 pt}
\begin{align}\label{eq:relay_3}
\notag        \vb*{y}_2 &= \vb*{H}_2 \ \vb*{x}_r + \vb*{n}_2,\\[-5pt]
                        &= \vb*{H}_2 \, \Bigg( \vb*{W}_r \, \bigg( \vb*{H}_1 \Big( \sum\limits_{k=1}^{K_r} \vb*{w}_{b,k} \ s_k \Big) + \, \vb*{n}_1   \bigg)  \Bigg) + \vb*{n}_2,\\[-20pt]
               \notag
\end{align}
where $\vb*{y}_2 \in \mathbb{C}^{K_r\times 1}$, $\vb*{H}_2 \in \mathbb{C}^{K_r\times N}$ is the A2G communication channel between the AR and the $K_r$ UEs, and $\vb*{n}_2 \sim \mathcal{CN}(0, \vb*{I}) \in \mathbb{C}^{K_r\times 1}$ is the noise vector. 
The $k$th user receives 
\vspace{-14 pt}
\begin{align}\label{eq:relay_4}
\notag         y_{2,k} &= \vb*{h}_{2,k} \vb*{W}_r \ \vb*{H}_1 \ \vb*{w}_{b,k} \ s_k  \\[-6pt]
\notag                 & + \vb*{h}_{2,k} \vb*{W}_r \ \vb*{H}_1 \Big( \sum\limits_{i\neq k}^{K_r} \vb*{w}_{b,i} \ s_i \Big)\\[-6pt] 
                        & +  \vb*{h}_{2,k} \vb*{W}_r \vb*{n}_1 + n_{2,k},\\[-20pt]
               \notag
\end{align}
where $\vb*{h}_{2,k} \in \mathbb{C}^{1\times N}$ denotes the MISO channel from the AR to the $k$th UE and $n_{2,k}\sim \mathcal{CN}(0, 1)$ is the additive noise. 
The SINR 
of this relayed communication link from the BS to the $k$th UE via the AR 
can then be calculated as
\vspace{-4 pt}
\begin{align}\label{eq:relay_5}
    \gamma_{r,k} &=  \frac{\mid  \vb*{h}_{2,k} \vb*{W}_r  \vb*{H}_1 \vb*{w}_{b,k}  \mid^2}{\sum\limits_{i\neq k}\mid  \vb*{h}_{2,k} \vb*{W}_r  \vb*{H}_1 \vb*{w}_{b,i} \mid^2 + \Vert \vb*{h}_{2,k} \vb*{W}_r \Vert^2 + 1 }\,.\\[-20pt]
               \notag
\end{align}
The channel capacity $C_{r,k}$ of the indirect link is obtained from (\ref{eq:direct_4}) using $\gamma_{r,k}$ instead of $\gamma_{b,k}$. \textcolor{black}{Note that $\vb*{H}_1$ and $\vb*{H}_2$ are directly influenced by UAV mobility due to changes in distance, altitude, and orientation relative to ground receivers.  }
\color{black}
\subsection{\color{black}Communication Model for Eavesdroppers}


\subsubsection{Eavesdropping on the direct communication link} The eavesdropper 
listens on the the direct link between the GBS and the \textcolor{black}{associated} UEs and receives
\vspace{-8 pt}
\begin{align}\label{eq:eaves_1}
\notag    y_{0,e} &=  \vb*{h}_{0,e} \ \vb*{x}_b + n_{e},\\[-4pt]
\notag            &=  \vb*{h}_{0,e} \ \big( \sum\limits_{k=1}^K \vb*{w}_{b,k} \, s_k \big) + n_{e},\\[-10pt]
                  &=  \vb*{h}_{0,e} \vb*{w}_{b,k}\,s_k \, + \vb*{h}_{0,e} \Big( \sum\limits_{\substack{i=1 \\ i\neq k}}^K  \vb*{w}_{b,i} \, s_i \Big) + n_{e},\\[-20pt]
               \notag
\end{align}
where $\vb*{h}_{0,e} \in \mathbb{C}^{1\times M}$ is the G2G communication channel between the BS and the eavesdropper and $n_{e}$ is the noise at the eavesdropper such that $n_{e}\sim \mathcal{CN}(0, 1)$. The SINR associated with the direct link between the GBS and the eavesdropper\textcolor{black}{---for the beam formed to user $k$---}can be calculated as
\vspace{-4 pt}
\begin{align}\label{eq:eaves_2}
    \textcolor{black}{\gamma_{b,e,k}}&=  \frac{\mid \vb*{h}_{0,e} \vb*{w}_{b,k} \mid^2}{\sum\limits_{{i\neq k}}\mid \vb*{h}_{0,e} \vb*{w}_{b,i} \mid^2 + 1} \,.\\[-20pt]
               \notag
\end{align}
Consequently, the capacity of the eavesdropper associated with the direct link from the BS to \textcolor{black}{the $k^{th}$ user} can be derived as
\vspace{-6 pt}
\begin{align}\label{eq:eaves_3}
\notag   \textcolor{black}{C_{b,e,k}} &= \log_2 \big(1+\gamma_{b,e,k}\big),\\[-1pt]
                &= \log_2 \bigg(1+\frac{\mid \vb*{h}_{0,e} \vb*{w}_{b,k} \mid^2}{\sum\limits_{{i\neq k}}\mid \vb*{h}_{0,e} \vb*{w}_{b,i} \mid^2 + 1 }\bigg).\\[-20pt]
               \notag
\end{align}


\subsubsection{Eavesdropping from relay communication link}
The eavesdropper can wiretap the A2G relay communication link between the UAV and the UEs. Similar to the section \ref{subsec:AR_model}, the capacity of the eavesdropper associated with the relay link can be derived as 
\vspace{-6 pt}
\begin{align}\label{eq:eaves_4}
\notag & C_{r,e} = \log_2 (1+\gamma_{r,e}) \\[-3pt]
&= \log_2 \bigg(1+ \frac{\mid  \vb*{h}_{2,e} \vb*{W}_r  \vb*{H}_1 \vb*{w}_{b,k}  \mid^2}{\sum\limits_{i\neq k}\mid  \vb*{h}_{2,e} \vb*{W}_r  \vb*{H}_1 \vb*{w}_{b,i} \mid^2 + \Vert \vb*{h}_{2,e} \vb*{W}_r \Vert^2 + 1 } \bigg),\\[-22pt]
               \notag
\end{align}
where $\gamma_{r,e}$ is the SINR, and $\vb*{h}_{2,e} \in \mathbb{C}^{1\times N}$ denotes the A2G channel between the UAV and the eavesdropper, and $n_{2,e}$ is the noise at the eavesdropper such that $n_{2,e}\sim \mathcal{CN}(0, 1)$. 

\color{black}

\subsection{\color{black}Secrecy Capacity}
\textcolor{black}{The term \textit{secrecy capacity} 
is a measure of the information rate that can be transmitted securely without being intercepted. It is 
obtained as the difference between the achievable data rate of a legitimate receiver and the achievable data rate of an eavesdropper, taking into account the channel conditions and the employed security measures. 
It corresponds to the rate at which no data will be decoded by the eavesdropper~\cite{YanWanGer2015}.  }
\color{black}

\color{black} For the system model of Section \ref{sec:system}, the average sum-secrecy capacity of the $K_b$ UEs that are directly served by the GBS over $T$ time slots is
\vspace{-5 pt}
\begin{align}\label{eq:secrecy_1}
\notag    {C_{sec,b}} &= \frac{1}{T}\sum\limits_{t=1}^T \sum\limits_{k=1}^{K_b} \Big( C_{b,k} \color{black}-  C_{b,e} \color{black} \Big)^+\\[-3pt]
\notag    &= \frac{1}{T}\sum\limits_{t=1}^T \sum\limits_{k=1}^{K_b} \Bigg[ \log_2 \bigg(1+\frac{\mid \vb*{h}_{0,k} \vb*{w}_{b,k} \mid^2}{\sum\limits_{{i\neq k}}\mid \vb*{h}_{0,k} \vb*{w}_{b,i} \mid^2 + 1 }\bigg) -\\[-6pt]
    & \qquad \qquad \quad \log_2 \bigg(1+\frac{\mid \vb*{h}_{0,e} \vb*{w}_{b,k} \mid^2}{\sum\limits_{{i\neq k}}\mid \vb*{h}_{0,e} \vb*{w}_{b,i} \mid^2 + 1 } \bigg) \Bigg]^+. \\[-22pt]
               \notag
\end{align}
Likewise, the average sum-secrecy capacity of the $K_r$ UEs served via the AR over $T$ time slots is obtained as
\small
\vspace{-5pt}
\begin{align}
\notag    {C_{sec,r}} &= \frac{1}{T}\sum\limits_{t=1}^T \sum\limits_{k=1}^{K_r} \Big( C_{r, k} \color{black}-  C_{r,e} \color{black} \Big)^+ = \frac{1}{T}\sum\limits_{t=1}^T\\[-27pt]
               \notag
\end{align}
\begin{align}\label{eq:secrecy_2}
\notag                & \sum\limits_{k=1}^{K_r} \Bigg[ \log_2 \bigg(1+ \frac{\mid  \vb*{h}_{2,k} \vb*{W}_r  \vb*{H}_1 \vb*{w}_{b,k}  \mid^2}{\sum\limits_{i\neq k}\mid  \vb*{h}_{2,k} \vb*{W}_r  \vb*{H}_1 \vb*{w}_{b,i} \mid^2 + \Vert \vb*{h}_{2,k} \vb*{W}_r \Vert^2 + 1 } \bigg) \\[-8pt]
    & - \log_2 \bigg(1+ \frac{\mid  \vb*{h}_{2,e} \vb*{W}_r  \vb*{H}_1 \vb*{w}_{b,k}  \mid^2}{\sum\limits_{i\neq k}\mid  \vb*{h}_{2,e} \vb*{W}_r  \vb*{H}_1 \vb*{w}_{b,i} \mid^2 + \Vert \vb*{h}_{2,e} \vb*{W}_r \Vert^2 + 1 } \bigg) \Bigg]^+.\\[-22pt]
               \notag
\end{align}
\normalsize
\color{black}
where $[\omega]^+ \triangleq max(\omega,0)$. 
The total secrecy capacity is
\vspace{-5pt}
\begin{equation}
C_T = C_{sec,b} + C_{sec,r}.
\vspace{-6pt}
\label{eq:total_sec}
\end{equation}

\textcolor{black}{Formulas (22)-(24) are derived from information theory and provide quantitative measures of the level of secrecy achieved in the communications channels according to the system model of Fig. 1. The secrecy capacity is maximized when maximizing the SINRs at the legitimate receivers and minimizing the SINRs at the eavesdroppers.}  

\vspace{-1mm}
\section{\color{black}
Problem Formulation} 
\vspace{-1 mm}
\label{sec:problem_def}
\color{black}
\color{black}





\color{black} 
This paper aims to maximize the 
total secrecy capacity of the UEs whether they are directly served by the GBS or through the AR. Considering the degrees of freedom for serving UEs directly or via the AR, we formulate two optimization problems.




\subsubsection{Direct communication}
For the directly served UEs
, the optimization problem is defined as 
\vspace{-4pt}
\begin{equation}\label{eq:optmzn_1}
\begin{aligned}
& \hspace{40pt}\underset{\vb*{w}_{b,k} }{\text{
max
}}
&& {C_{sec,b}} \\[-4pt]
& \text{subject to (s.t.)}  && P_b \leq P_{b, max},
\end{aligned}
\end{equation}
where $C_{sec,b}$ is the secrecy capacity 
defined in (\ref{eq:secrecy_1}), $P_{b,max}$ is the maximum transmit power of the GBS, and $P_{b}$ is the transmit power of the GBS, that is, $P_b = $ Tr \big($ \vb*{x}_b \vb*{x}_b^{\dagger}$\big).

Problem (\ref{eq:optmzn_1}) requires the knowledge of the eavesdropping channel. We assume the location of the eavesdropper and thus its CSI  to be unknown, which is the scenario of interest in practice where it is difficult to detect or estimate the presence, location, or channel of eavesdroppers because of their passive nature. 
Therefore, we can only consider the capacity of the legitimate user and reformulate the optimization problem: 
\vspace{-5pt}
\begin{equation}\label{eq:optmzn_2}
\begin{aligned}
& \underset{\vb*{w}_{b,k} }{\text{max}}
&& \frac{1}{T}\sum\limits_{t=1}^T \sum\limits_{k=1}^K \Bigg[ \log_2 \bigg(1+\frac{\mid \vb*{h}_{0,k} \vb*{w}_{b,k} \mid^2}{\sum\limits_{{i\neq k}}\mid \vb*{h}_{0,k} \vb*{w}_{b,i} \mid^2 + 1 }\bigg) \Bigg] \\[-7pt]
& \text{s.t.}  && \text{Tr} \big( \vb*{x}_b \vb*{x}_b^{\dagger} \big) \leq P_{b, max} .\\[-5pt]
\end{aligned}
\end{equation}

The eavesdropper location and channel are used only for calculating the resulting secrecy capacity 
for performance evaluation.

\subsubsection{Relay communication}
For the UEs that are served 
via the AR, the optimization problem 
is defined as
\vspace{-6pt}
\begin{equation}\label{eq:optmzn_3}
\begin{aligned}
& \underset{\{\vb*{w}_{b,k}, x_r, y_r, z_r\}}{\text{max}}
&& {C_{sec,r}} \\[-6pt]
& \text{s.t.}  && P_{r} \leq P_{r, max} \\[-5pt]
&&& (x_r,y_r, z_r) \leq (L_x,L_y, L_z),
\end{aligned}
\end{equation}
where $C_{sec,r}$ is the secrecy capacity 
defined in (\ref{eq:secrecy_2}), $P_{r,max}$ is the maximum transmit power of the AR, $P_r = $ Tr \big($ \vb*{x}_r \vb*{x}_r^{\dagger}$\big) is the transmit power of the AR and $P_{r,max}$ the maximum transmit power, 
and \color{black} $(x_r,y_r, z_r)$ are the 3D coordinates of the UAV bound to $(L_x,L_y, L_z)$. 
Because of the unknown eavesdropper location and CSI, the optimization problem is rewritten as 
\vspace{-5pt}
\begin{equation}\label{eq:optmzn_4}
\begin{aligned}
& \underset{\{\vb*{W}_{r}, x_r, y_r, z_r\}}{\text{max}}
 && \frac{1}{T}\sum\limits_{t=1}^T \sum\limits_{k=1}^K \Bigg[ \log_2 \bigg(1 + \\[-3pt]
 &&& \frac{\mid  \vb*{h}_{2,k} \vb*{W}_r  \vb*{H}_1 \vb*{w}_{b,k}  \mid^2}{\sum\limits_{i\neq k}\mid  \vb*{h}_{2,k} \vb*{W}_r  \vb*{H}_1 \vb*{w}_{b,i} \mid^2 + \Vert \vb*{h}_{2,k} \vb*{W}_r \Vert^2 + 1 } \bigg) \Bigg] \\[-4pt]
& \text{s.t.}  &&  \text{Tr} \big( \vb*{x}_r \vb*{x}_r^{\dagger} \big) \leq P_{r, max} \\[-3pt]
&&& (x_r,y_r, z_r) \leq (L_x,L_y, L_z).\\[-5pt]
\end{aligned}
\end{equation}


\textcolor{black}{Although we have incorporated practical system constraints in our model, we acknowledge that there are additional operational aspects, such as UAV energy consumption, flight time, and speed~\cite{Reviewerpaper}, which are not optimized in this paper.}

\section{\color{black} Proposed Solution}
\vspace{-1 mm}
\label{sec:solution}


\textcolor{black}{Given the available resources, which are one multi-antenna GBS and one multi-antenna AR, 
the secrecy capacity 
optimization problem becomes a user association and transmission parameter optimization problem. 
}
\begin{figure*}[ht]
	   \centering
	   \includegraphics[width=0.85\textwidth]{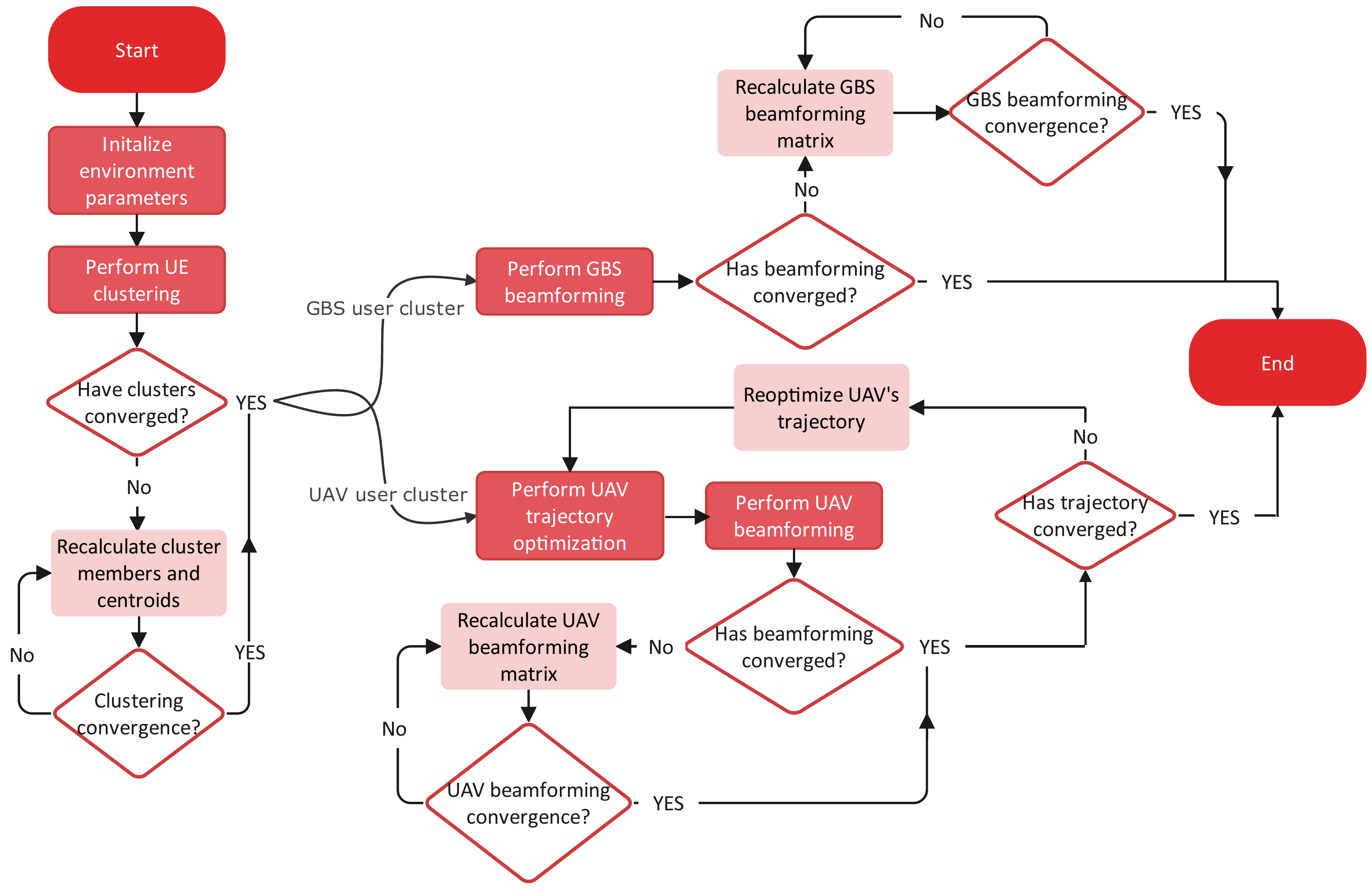}
	    \vspace{-9 pt}
        \caption{Proposed solution flowchart. 
        }
        \label{fig:Flowchart}
\end{figure*}
We perform UE clustering for user association, followed by GBS and AR beamforming and transmit power control, and UAV trajectory optimization. Figure~\ref{fig:Flowchart} illustrates this.
\color{black} 
It is important to mention that the beamforming/power control and the UAV trajectory optimization are done through an iterative process. That is, the algorithm obtains the optimal power coefficients for every 3D location of the UAV. Hence, the beamforming and transmit power control of the UAV affects its trajectory adjustment. The details are discussed in Sections IV.B and IV.C
\color{black}



\color{black}

\subsection{User Clustering} 
\vspace{-1 mm}
\label{sec:cluster}
\color{black}
The goal of user clustering is to divide $K$ users into two clusters, one cluster is to be served by the GBS and the other cluster is served the UAV.

For solving the user clustering problem, we can employ an exhaustive search, but it entails a high computational complexity, which increases exponentially with the number of users. We instead apply K-means clustering, a unsupervised machine learning algorithm that is used for grouping a set of objects so that the similarity criterion of members in a group and the dissimilarity with members of other groups is maximized. 
K-mean works with any single or multi-dimensional metric that the data captures and user-defined target number of clusters \cite{VukVTC17}. It is a computationally efficient method  compared to other techniques such as graph theory, fuzzy c-means clustering, and hierarchical clustering~\cite{clustering}. 

\color{black}
We consider the characteristics of wireless communication systems to determine the similarities of the data points. Because the objective it to associate users to base stations, one fixed (GBS) and one mobile (UAV), 
we take the normalized channel coefficients between UEs and GBS as the data points of the K-means clustering algorithm. This captures the variations of channel gains resulting from different RF propagation effects such as small-scale fading and shadow fading. Hence, we can define
\vspace{-5 pt}
\begin{equation}
        h^n_{b,k} = \frac{ h_{b,k}}{\parallel h_{b,k} \parallel_2},
    \label{eq:sim}
\vspace{-4pt}
\end{equation}
where $h^n_{b,k}$ is the normalized channel gain, $h_{b,k}$ is the channel gain between the GBS and $k$-th UE, and $\parallel.\parallel_2$ is the $L_2$ vector norm. Having these channel gains as data points, we apply K-mean clustering algorithm to determine the cluster centers, or centroids, and consequently the UEs associated with each centroid. The goal is to leverage the similarity of channels between 
the GBS and the UEs to create two UE clusters, where the UEs of one cluster are to be served by the GBS and the UEs of the other cluster 
by the UAV. \textcolor{black}{This approach is applying the same clustering principle as other studies in the literature~\cite{ChannelClustering1, ChannelClustering2, ChannelClustering3, ChannelClustering4}.} 

The K-mean clustering algorithm can be done as follows \cite{alpaydin2020ML, geron2022hands}: (i) the initial centroids $C=\{c_1, c_2, \dots c_n \}$ are randomly selected as the $n$ cluster centers of the $K$ available data points: $U=\{u_1, u_2, \dots , u_k \dots, u_K\}$. \color{black} Here, we consider two clusters $c_1$ and $c_2$ for the GBS and the UAV, and $K$ users where $u_k = h^n_{b,k}$ can be considered as a data point of the $k$-th user. 
(ii) Distance between each data point, e.g., channel status, and the cluster centers is calculated to assign the data point to the nearest center. \color{black} Different metrics can be used to measure the distance between data points such as Euclidean distance, Manhattan distance, etc. In this paper, we use $L^{2}_2$ norm or Euclidean distance. (iii) the centroids are updated to minimize the sum of squared distances between a user and its centroid,
\vspace{-5pt}
\begin{equation}
    \min\limits_{C} \sum \limits_{k} \min\limits_{r\in R} \ d_{r,k} 
    \label{eq:K-meansCenterUpdate}
    \vspace{-7pt}
\end{equation}
where $d_{r,k} = \ \parallel u_{k} - c_r\parallel^2_2$ is the Euclidean distance between $C \triangleq \{c_r \mid r \in R\}$ and $R=2$ represents the number of clusters. For example, the distance between the normalized channel gains and the centroids is $d_{r,k}  = \ \parallel h^n_{b,k} - c_r  \parallel^2_2$. Algorithm 1 represents a pseudocode for the K-mean clustering algorithm. The implementation of algorithm 1 for one scenario is shown in Figure~\ref{fig:Clustering} wherein data points are $h^{n}_{b,k}$

\color{black} In addition to UEs' channel status, other data points can also be considered in the algorithm to study the problem. \color{black}For example, distance based clustering or rate based clustering, where UEs with the nearest distance to the GBS and the highest downlink rates, respectively, would be grouped. 
Each scenario can have an effect on the system performance and should be chosen based on the objective and the ability or simplicity to obtain the necessary information to calculate the value for each UE. \color{black} In Section V.C, we discuss about the results of different scenarios.
\color{black} 

\begin{figure}[ht]
	   \centering
	   \includegraphics[width=0.45\textwidth]{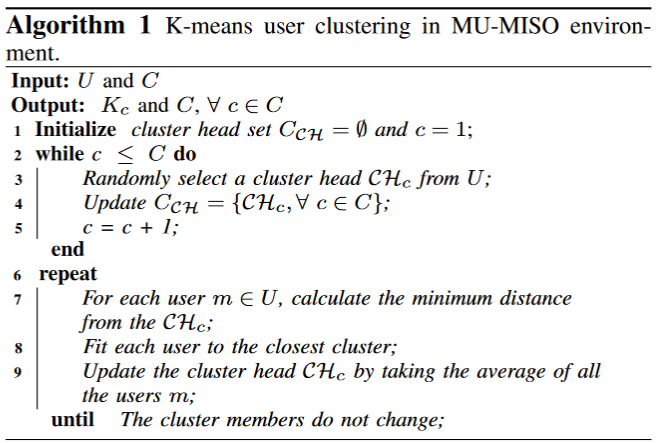}
\end{figure}

\begin{figure}[ht]
	   \centering
	   \includegraphics[width=0.45\textwidth]{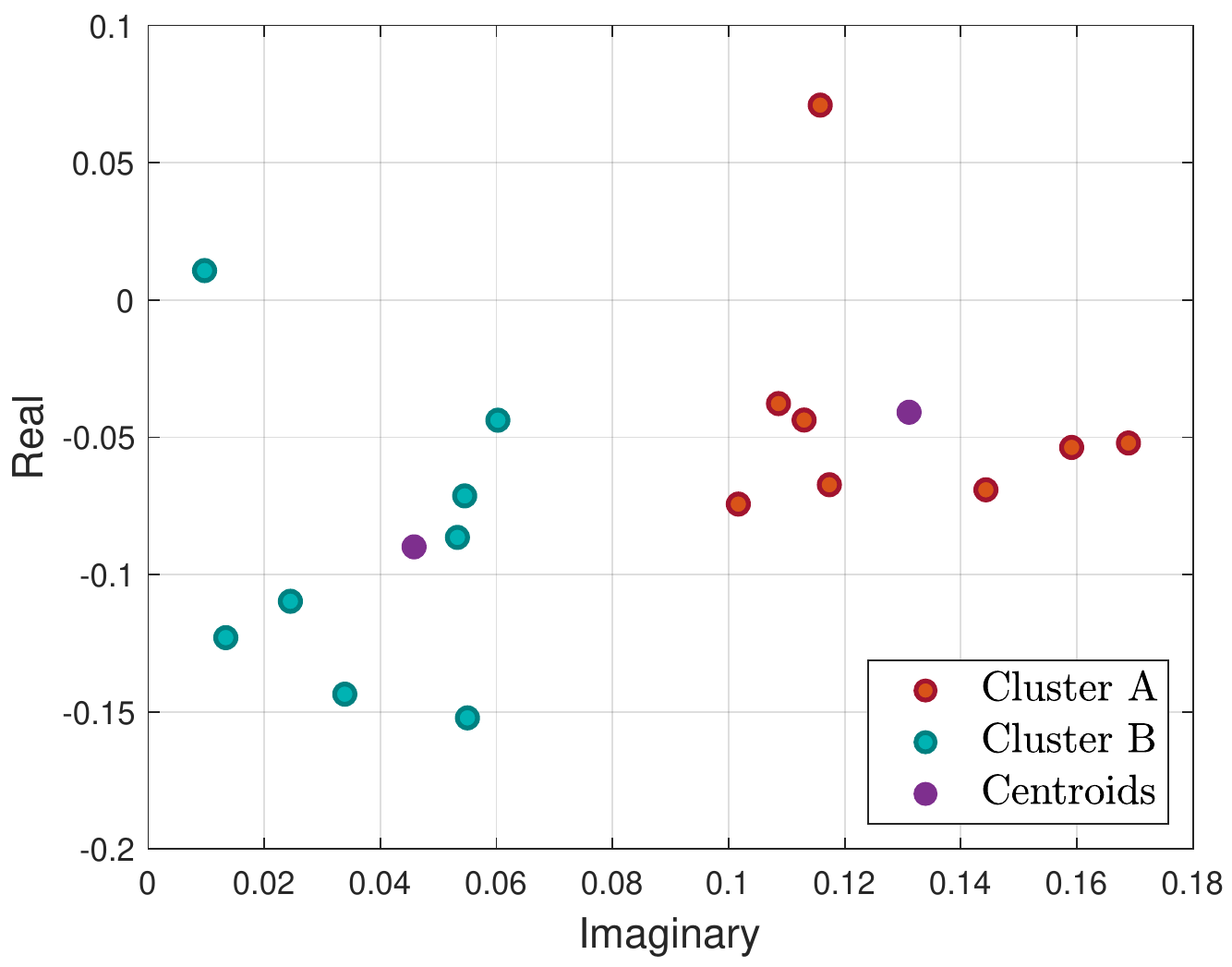}
        \caption{Channel-based clustering. 
        }
        \label{fig:Clustering}
\end{figure}

\color{black}

\subsection{\color{black} Optimal 
Beamforming and Power Control}
\vspace{-1 mm}
\label{sec:optimal_power}
\color{black}
Problem (\ref{eq:optmzn_2}) is a traditional beamforming power control problem between a GBS and a user. This problem has been extensively studied in the literature \cite{WMMSE1, WMMSE2}. \textcolor{black}{The beamforming vectors of the GBS are obtained by applying the weighted minimum mean square error (WMMSE) algorithm~\cite{WMMSE}. The WMMSE is an iterative closed form solution 
that optimizes the transmitter and receiver precoding vectors to maximize the sum rate of all UEs for a GBS power constraint. The precoding solution for the direct communication links is then~\cite{WMMSEs}
\vspace{-5pt}
\begin{equation}
    \vb*{W_{bk}}=\Big(\vb*{H_0^H}\vb*{Q^H}\vb*{F}\vb*{Q}\vb*{H_0}+\frac{\text{Tr}(\vb*{F}\vb*{Q}\vb*{Q^H})}{P_{b, max}}\vb*{I_M}\Big)^{-1} \vb*{H_0^H}\vb*{Q^H}\vb*{F}
    \vspace{-2pt}
\end{equation}
where $Q=diag\{q_1,\cdots,q_k\}$ is the receiver precoding, $F=diag\{f_1,\cdots,f_k\}$ is the  weight matrix, and $\vb*{I_M}$ is the covariance matrix.} 

The beamforming and power control for the relay communications problem is solved in the remainder of this section. 
Inspired by the zero-forcing (ZF) criterion and the channel singular value decomposition (SVD) based structure introduced in \cite{zhang09}, we first propose a beamforming matrix structure for the UAV (i.e., $\vb*{W}_r$) to eliminate interference among users. This converts the optimization problem (\ref{eq:optmzn_4}) into a simplified convex optimization problem. Then, we solve the modified optimization problem using the Lagrangian function and Karush-Kuhn-Tucker (KKT) conditions to obtain the UAV's optimal beamforming matrix.


\subsubsection{
Beamforming Matrices}
Beamforming is done at the GBS and the AR, each serving a distinct set of users. By using the concepts of channel inversion, ZF, and linear algebra, the multi-user interference 
can be minimized. From (\ref{eq:relay_3}), the received signal at $K$ UEs transmitted from the UAV can be written as
\vspace{-8pt}
\begin{align}\label{eq:pwr_cntrl_1}
        \vb*{y}_2 &= \vb*{H}_2 \vb*{W}_r \vb*{H}_1 \vb*{W}_{br} \, \vb*{s}_K \, + \,
              \vb*{H}_2 \vb*{W}_r \, \vb*{n}_1 \, + \, \vb*{n}_2,\\[-19pt]
               \notag
\end{align}
where $\vb*{s}_K \in \mathbb{C}^{K \times 1}$ corresponds to the $K$ transmit signals to the $K$ UEs. The ZF criterion requires that $ \vb*{H}_2 \vb*{W}_r \vb*{H}_1 \vb*{W}_{br}$ is to be a diagonal matrix with rank $K$, which implies that Rank$(\vb*{H}_1) \geq K$ and Rank$(\vb*{H}_2) \geq K$ \cite{zhang09}. By applying the SVD, $\vb*{H}_2$ and $\vb*{H}_1$ can be expressed as
\vspace{-4pt}
\begin{align}\label{eq:pwr_cntrl_2}
        \vb*{H}_1 &= \vb*{U}_1 \, \vb*{\Sigma}_1 \, \vb*{V}_1 \ssymbol{2},\\[-17pt]
               \notag
\end{align}
\vspace{-0.8cm}
\begin{align}\label{eq:pwr_cntrl_3}
        \vb*{H}_2 &= \vb*{U}_2 \, \vb*{\Sigma}_2 \, \vb*{V}_2 \ssymbol{2},
\end{align}
where $\vb*{U}_i$ and $\vb*{V}_i$, for $i= 1, 2$, are unitary matrices and $\vb*{\Sigma}_i \in \mathbb{C}^{K \times K}$ is a diagonal matrix with positive diagonal elements. Knowing the channel coefficients at the BS and at the UAV, 
to satisfy the ZF criterion, we propose the following beamforming matrices for the GBS and the UAV
\vspace{-6pt}
\begin{align}\label{eq:pwr_cntrl_4}
        \vb*{W}_{br} &= \vb*{V}_1 \, \vb*{\Lambda}_b \, \vb*{U}_2 \ssymbol{2},
\end{align}
\vspace{-0.8cm}
\begin{align}\label{eq:pwr_cntrl_5}
        \vb*{W}_r &= \vb*{V}_2 \, \hat{\vb*{\Lambda}}_r \, \vb*{U}_1 \ssymbol{2} \, \vb*{\Lambda}_r,\\[-20pt]
               \notag
\end{align}
where $\vb*{\Lambda}_b, \hat{\vb*{\Lambda}}_r$, and $\vb*{\Lambda}_r$ are all $K \times K$ diagonal matrices. Without loss of generality, it can be assumed that the elements of these two diagonal matrices are non-negative, representing the allocated beamforming power at the BS and the UAV, respectively. 

\subsubsection{Optimal AR Transmit Power Allocation}
\begin{lemma}\label{lem:objective}
The objective function defined in (\ref{eq:optmzn_4}) can be written as
\vspace{-10pt}
\begin{align}\label{eq:pwr_cntrl_10}
        \frac{1}{T} \sum\limits_{t=1}^T \sum\limits_{k=1}^K log_2 \Big( 1+ \frac{\lambda_{r,k}^2}{\lambda_{r,k}^2+1}\Big),\\[-19pt]
               \notag
\end{align}
where $\lambda_{r,k}$ is the $\vb*{\Lambda}_r(k,k)$.
\end{lemma}

\begin{lemma}\label{lem:const}
The beamforming power constraint defined in (\ref{eq:optmzn_4}) can be expressed as
\vspace{-7pt}
\begin{align}\label{eq:pwr_cntrl_11}
        2 \, \sum\limits_{m=1}^K  \sum\limits_{n=1}^K |\vb*{U}_2(m,n)|^2 \, \sigma_{2,n}^{-2} \, \lambda_{r,m}^2 \leq P_{r, max},\\[-20pt]
               \notag
\end{align}
where $\sigma_{2,n}$ is the $\vb*{\Sigma}_2(n,n)$.
\end{lemma}
The lemmas 
are proved in the Appendix (Section \ref{sec:Appendix}). 
Leveraging (\ref{eq:pwr_cntrl_10}) and (\ref{eq:pwr_cntrl_11}), the beamforming power optimization problem for the UAV at any location can be written as 
\vspace{-15pt}
\begin{align}
& \underset{\{\lambda_{r,m}\}}{\text{max}}
 && \frac{1}{T} \sum\limits_{t=1}^T \sum\limits_{m=1}^K log_2 \Big( 1+ \frac{\lambda_{r,m}^2}{\lambda_{r,m}^2+1}\Big) \label{eq:pwr_cntrl_12}\\[-3pt]
& \text{s.t.}  &&  2 \, \sum\limits_{m=1}^K  \sum\limits_{n=1}^K |\vb*{U}_2(m,n)|^2 \, \sigma_{2,n}^{-2} \, \lambda_{r,m}^2 \leq P_{r, max} \label{eq:pwr_cntrl_12_1},\\[-3pt]
&&& 0 \leq \lambda_{r,m} \leq \lambda_{r, max}, \quad m \in \{1, ...,K  \},\\[-20pt]
               \notag              
\end{align}
where $P_{r, max}$ is the maximum available transmit power at the UAV, and $\lambda_{r, max}$ is the maximum allocated power for each 
antenna.

The Lagrangian function of the optimization problem can be expressed as 
\vspace{-12 pt}
\begin{align}\label{eq:pwr_cntrl_13}
\notag &\mathcal{L}(\lambda_{r,l}, \alpha_1, \alpha_{2,l}, \alpha_{3,l}) = + \sum\limits_{l=1}^K log_2 \Big( 1+ \frac{\lambda_{r,l}^2}{\lambda_{r,l}^2+1}\Big)\\[-10pt]
\notag  & - \alpha_1 \bigg( 2 \, \sum\limits_{l=1}^K  \sum\limits_{n=1}^K |\vb*{U}_2(l,n)|^2 \, \sigma_{2,n}^{-2} \, \lambda_{r,l}^2 - P_{r, max}  \bigg) \\[-4pt]
&- \bigg(\alpha_{2,l} \Big(\sum\limits_{l=1}^K  \lambda_{r,l} - \lambda_{r, max} \Big) \bigg) - \bigg(\alpha_{3,l} \sum\limits_{l=1}^K  - \lambda_{r,l} \bigg),\\[-19pt]
               \notag
\end{align}
where $\alpha_1$, $\alpha_{2,l}$, and $\alpha_{3,l}$ are the non-negative Lagrangian multipliers corresponding to the first and the second constraints, respectively. 

\begin{theorem}\label{thm:optimization_1}
The optimal beamforming power for the $l$th antenna of the UAV can be obtained as
\vspace{-4pt}
\begin{equation*}
\lambda_{r,l}^{*}=
\begin{cases}
\small          0 \qquad    &\quad  \lambda_{r,l}^{\dagger} \leq 0, (\alpha_1^{*} \, F\,  \text{ln}\,2)>0.25 \\
          \lambda_{r,l}^{\dagger} \qquad   &\qquad 0 < \lambda_{r,l}^{\dagger} < \lambda_{r,max}\\
          \lambda_{r,max} \qquad  &\qquad \quad \ \ \lambda_{r,l}^{\dagger} \geq \lambda_{r,max}
\end{cases}
\end{equation*}
in which
\vspace{-4pt}
\begin{align} \label{eq:pwr_cntrl_19}
   \lambda_{r,l}^{\dagger} = \sqrt{\frac{1}{4}\Big(\sqrt{1+\frac{2}{\alpha_1^{*} \, F \, \text{ln}\,2 }} - 3\Big)},\\[-22pt]
               \notag
\end{align}
where $F$ is a constant and equals to $\sum\limits_{n=1}^K |\vb*{U}_2(l,n)|^2 \, \sigma_{2,n}^{-2}$, and the Lagrangian multiplier $\alpha_1^{*}$ can be obtained by replacing (\ref{eq:pwr_cntrl_19}) into the first constraint of (\ref{eq:pwr_cntrl_12_1}) when the equality holds.
\end{theorem}

The optimal beamforming matrix and transmit power formulations for the AR 
defined above 
are used for the UAV trajectory optimization.



\color{black}

\subsection{UAV Trajectory Optimization
} 
\vspace{-1 mm}
\label{sec:solution}


\textcolor{black}{The objective function of (\ref{eq:optmzn_4}) is non-convex with respect to parameters $x_r$, $y_r$, $z_r$, $P$, and the
constraints, and the problem is NP-hard \cite{TVT, MADRL,li2018deep,zhang09}. 
We, therefore, propose a machine learning solution } 
where the UAV trajectory 
is updated through a transition process based on the current system state. Since the next system state is independent from the previous state and action, the process can be modeled as a Markov decision process (MDP). 
In order to avoid intractably high dimensionality for the high state-action space, we 
propose a 
DQN. 
\color{black}
It is noteworthy that the following proposed DQN is based on basic reinforcement learning algorithms such as Q-learning and deep reinforcement learning. The aim is to use DQN as an alternative tool for solving this NP-hard optimization problem  while consuming less power and computational resources~\cite{DQL_QL,DQLEV,QLEV}. Depending on an application, one can extend the following framework to more advanced learning models that suit particular use cases. 
\color{black}

\subsubsection{MDP Settings}
The MDP for the UAV agent is composed of the state space $\mathcal{S}$, the action space $\mathcal{A}$, the reward space $\mathcal{R}$, and the transition probability space $\mathcal{T}$. 
At time slot $t$, the agent observes the state $s_t \in \mathcal{S}$, and takes action $a_t \in \mathcal{A}$ based on its policy. Depending on the distribution of the transition probability $\mathcal{T}(s_{t+1}|s_t, a_t)$, the agent is then transferred to the new state $s_{t+1}$. Since the transition probability is specific to the 
operational environment, we choose the Q-learning method as a model-free algorithm to find the best policy for each action in each state. This means that we do not need to know $\mathcal{T}$, but we need to carefully define the states, the actions, and the reward. 

\textbf{State:} The set of states is defined as $\mathcal{S} = \{s_1, s_2, ..., s_t, .., s_T\},$
where $t$ is the time slot index. 
Each state $s_t$ 
corresponds to the 3D coordinates of the UAV and 
the users served by the AR. 

\textbf{Action:} The states are transitioned according to the defined set of actions defined as $\mathcal{A} = \{a_1, a_2, ..., a_t, .., a_T\}$, 
where each action 
consists of three parts related to the UAV movement, 
$a_t = \{\delta_x, \delta_y, \delta_z\}$, where $\delta_x$ , $\delta_y$, and $\delta_z$ represent the movement in the $x$, $y$ 
and $z$ directions. 
The movement along each axis 
is assumed to change  positively or negatively, or remain in the original position. 
Hence, here we consider $3$ possible directional movements for the $3$ axes of the AR trajectory, 
resulting in $27$ possible actions for the AR. 


 \textbf{Reward:} After taking action $a_t$ in state $s_t$, 
 the UAV agent will receive a reward $R_t(s_t, a_t)$. The UAV gets more rewards for actions that lead to higher legitimate user rates. 
 We 
 define the reward function accordingly: 
 \vspace{-6pt}
\begin{equation}
    {R_t(s_t, a_t)} =  \sum\limits_{k=1}^{K_r} C_{r,k} 
    \label{eq:reward}
\end{equation} 



%
%
\subsubsection{Deep Q-Network Method}
The DQN
, initially proposed by Google Deep Mind \cite{DeepMind2015}, integrates the RL and deep learning methods. This technique uses the power of nonlinear functions, specifically DNNs, in order to approximate the Q-values and handle highly dimensional state-action problems. 

There are two DNNs of the same structure: a training network and a target network. The training network outputs the Q-values associated with the actions of the UAV in each state. The target network supervises the training network by providing the target Q-values obtained from the Bellman equation  \cite{Survey17deep}, 
\begin{flalign}
Q^{*}(s, a) = E_{s^\prime}\bigg[R(s, a) + \gamma \times \max_{a\in \mathcal{A}} \ Q(s^\prime, a^\prime)\bigg],
 \label{eq:Bellman}
 \end{flalign}
which provides the optimal state-action pairs, where $s^\prime$ and $a^\prime$ symbolize the next state and action. Parameter $\gamma \in (0,1)$ denotes the discount factor that affects the importance of the future reward.

The target values are compared with the outputs of the training network to minimize the loss function,
\begin{flalign}
\notag L(\theta) = \mathbb{E} \Bigg[ \bigg( \Big[ r_t + \gamma \times \max_{a\in \mathcal{A}} \ Q(s_{t+1}, a_{t+1}; \theta^{\dagger})\Big] - \\ 
\Big[Q(s_{t}, a_{t}; \theta) \Big] \bigg)^2 \Bigg],
\label{eq:loss}
\end{flalign}
where the Q-value of the first term is obtained from the target network and the Q-value of the second term 
is obtained from the training network. 
Parameters $\theta^{\dagger}$ and $\theta$ denote the weights of the target network and training network, respectively. The $\theta^{\dagger}$ coefficients are updated every few time slots in order to ensure the stability of the target values and, hence, facilitate stable learning. 

\begin{figure}[ht]
\vspace{10 pt}
	   \centering
	   \includegraphics[width=0.45\textwidth]{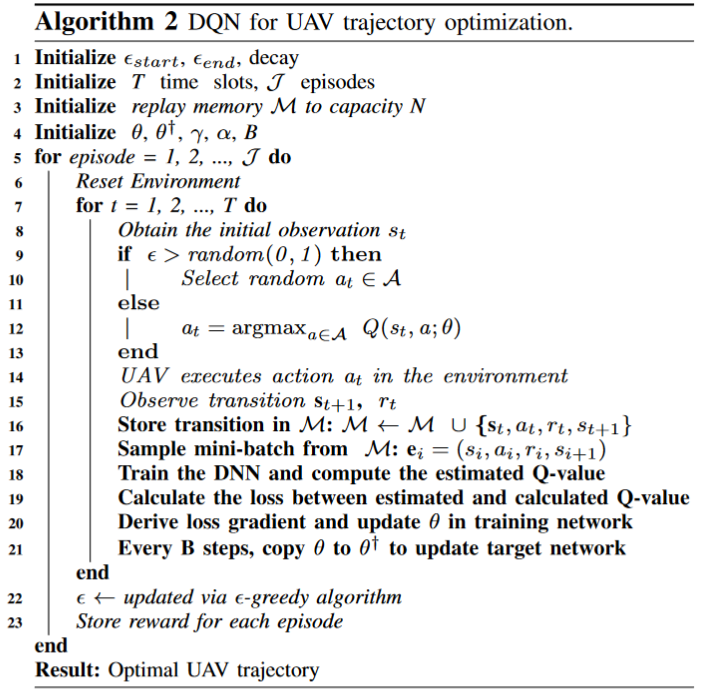}
\end{figure}

As the UAV takes an action
, the system generates a record of experience. At time step $t$, the experience contains the current state $s_t$, the action $a_t$, the reward $r_t$, and the next state $s_{t+1}$, formed as a tuple $e_t = (s_t, a_t, r_t, s_{t+1})$. Each such experience is stored in a replay memory with the capacity of $N$, such that $\mathcal{M}=\{e_1, ..., e_t, ..., e_N \}$. The memory is a queue-like buffer that stores the latest N experience vectors. 
We use a mini-batch sample from the replay memory to feed the input of the training network. 
The main reason for using the mini-batch samples from the reply memory is to break possible correlations between sequential states of the environment, and thereby facilitate generalization.


The UAV applies a gradient descent algorithm,
\begin{flalign}
\notag \nabla_{\theta}\, L(\theta) &= - \mathbb{E} \Bigg[ 2 \ \nabla_{\theta}Q(s_{t}, a_{t}; \theta) \bigg( \, r_t + \, \gamma \ \times \\ & \qquad \max_{a\in \mathcal{A}} \ Q(s_{t+1}, a_{t+1}; \theta^{\dagger})  - Q(s_{t}, a_{t}; \theta) \bigg)  \Bigg],
\label{eq:GDS}
\end{flalign}
to update $\theta$ an
$\theta^{\dagger}$ as the weights of the DNNs with the aim of minimizing the prediction error.

Finally, we apply the $\epsilon-$greedy algorithm to select an action while balancing the exploration and the exploitation of the UAV in the environment. 
In this algorithm, the UAV explores the environment with the probability of $\epsilon$ by choosing a random action. 
More precisely, the UAV exploits the environment with the probability of $1-\epsilon$ by choosing the actions that maximize the Q-value function, i.e., $a^{*} = \text{argmax}_{a \in \mathcal{A}} \ Q(s,a; \theta)$. A high value of $\epsilon$ is initially set in the model for the UAV to spend more time for the exploration. As the agent obtains more knowledge about the environment, the $\epsilon$ value is gradually decreased to leverage the experience and choose the best actions for the UAV, rather than continuing with the exploration. 

Algorithm 2 details the DQN-based algorithm used by the UAV agent for optimizing the sum-rate of the UEs that are served via the AR. 

\color{black}
In summary, the 
proposed techniques 
accomplish the following: 
i) User-BS association employing K-means clustering 
(e.g., channel, rate, or distance based), ii) multi-user beamforming and power management for the 
GBS, iii) UAV trajectory optimization in conjunction with multi-user beamforming and power management. 
The objective is to maximize the secrecy 
rate which can be used to evaluate the effectiveness of the proposed security measure
in protecting against eavesdropping attacks~\cite{SR1, SR2}, especially in the context of wireless communication~\cite{mamaghani2020intelligent, kang2019secrecy, wang2019optimal, zhang2020uav}. Since the CSI of the eavesdropping channel cannot be obtained for passive, receive-only eavesdroppers, our solution maximizes the user rate for it to be generally applicable without requiring collaboration with eavesdroppers or wasting power for generating artificial noise in random directions, because of the unknown eavesdropper locations, 
as opposed to using this power to increase the user rate. The secrecy capacity also provides a unified measurement framework for 
the numerical analyses presented in the following section. 
\color{black}

\section{Numerical Analysis and Discussion}
\vspace{-1 mm}
\label{sec:results}

In this section, we present simulation results to evaluate the secrecy performance of the 
UAV-assisted 
communications system, where users are clustered and served by a fixed and a mobile access point. 
In the presence of an eavesdropping attack, our solution jointly optimizes of the UAV trajectory, GBS beamforming, and AR beamforming coefficients. 
The 
numerical analysis quantifies the impact of different user clustering technique, discount factor (Gamma) values, learning rates, and number of users on the achievable secrecy capacity 
of the system. 

The simulation scenario is illustrated in Fig. 1 and consist of multiple single antenna ground UEs, an AR, and a group of malicious nodes that is performing a passive eavesdropping attack on the downlink transmission. The terrestrial users and the eavesdroppers are randomly distributed in a 2D area. 
The AR is launched at a random location and height and is equipped with an antenna array to enable communications with the GBS and the UEs. Table~II captures the simulation parameters. The simulations are performed with Python 3.6 and PyTorch 1.7. 

\begin{figure}[ht]
\vspace{10 pt}
	   \centering
	   \includegraphics[width=0.45\textwidth]{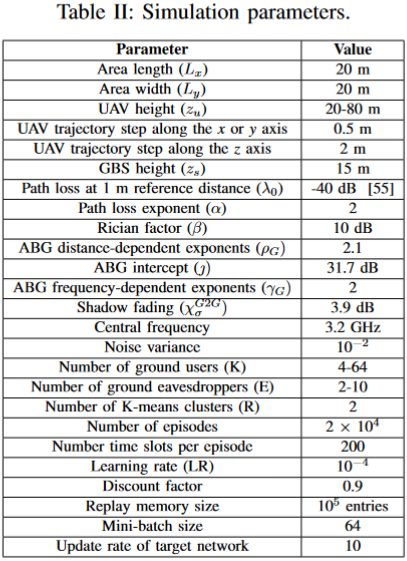}
\end{figure}

\subsection{
Hyper-parameters}

\begin{figure}[b]
\vspace{-7 mm }
	   \centering
	   \includegraphics[width=0.45\textwidth]{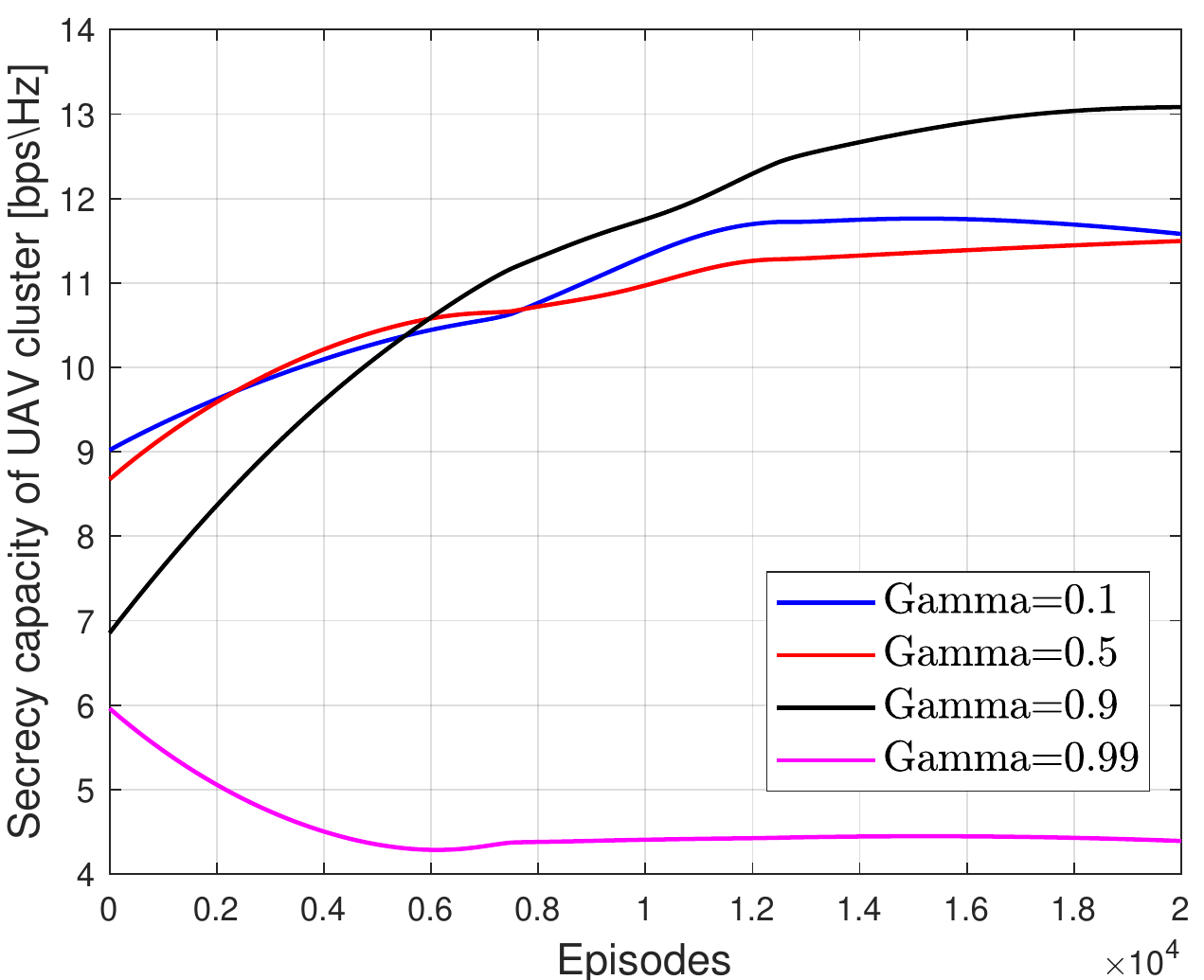}
	    \vspace{-8 pt}
        \caption{DQL performance for different discount factors Gamma for a learning rate of $10^{-4}$. 
        }
        \label{fig:GammaComp}
\end{figure}

\begin{figure}[ht]
	   \centering
	   \includegraphics[width=0.45\textwidth]{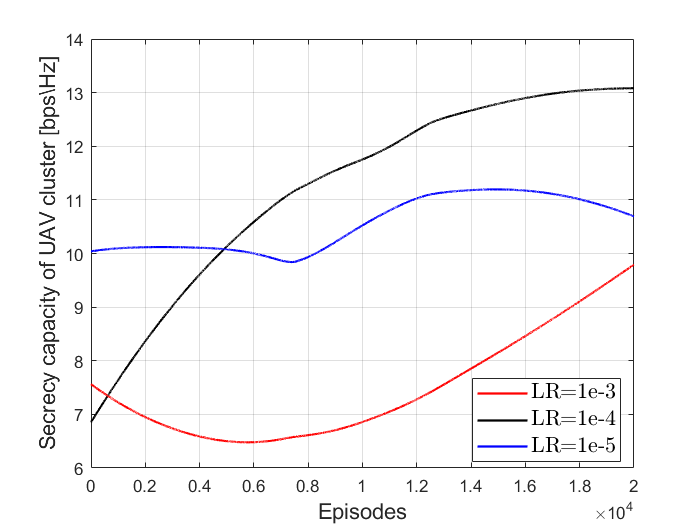}
	    \vspace{-8 pt}
        \caption{
        DQL performance for different learning rates (LRs) for Gamma = 0.9. 
        }
        \vspace{-5 mm}
        \label{fig:LRComp}
\end{figure}

The hyper-parameters of the learning algorithm need to be optimized for our specific problem and environment. Therefore,  Fig.~\ref{fig:GammaComp} and Fig.~\ref{fig:LRComp} numerically evaluate the secrecy capacity of the UEs served through the AR 
for different discount factors Gamma and learning rates  (LRs). \textcolor{black}{Additionally,  Fig.~\ref{fig:GammaComp} and Fig.~\ref{fig:LRComp} verify the convergence of the proposed solution across these different settings.} 
The results presented in both figures 
are for the case of 16 ground users where there are 8 users in each clusters as shown earlier in Fig.~\ref{fig:Clustering}. 
These figures plot the total achieved secrecy capacity 
of the user cluster served by the AR over the training time for different hyper-parameter values. 


\begin{figure}[ht]
\vspace{-3 pt}
	   \centering
	   \includegraphics[width=0.45\textwidth]{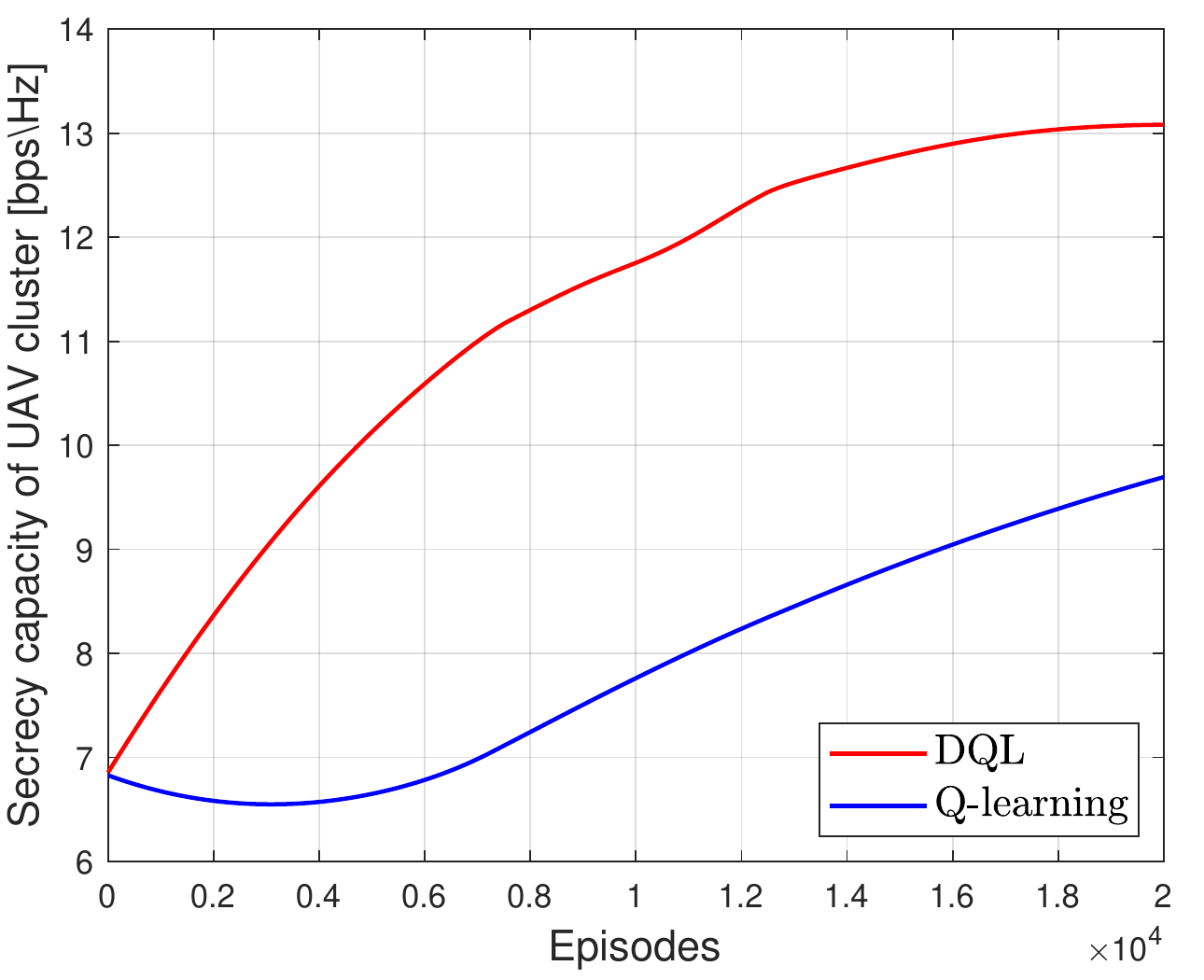}
	    \vspace{-8 pt}
        \caption{Comparison of the learning performance of the DQL and Q-learning for the UAV trajectory optimization.
        }
        \label{fig:DQL_QL}
\end{figure}

When the discount factor is very high, the agent equally considers the 
future and current rewards. 
Fig.~\ref{fig:GammaComp} shows that this leads to low performance. 
The best result for our scenario is achieved by slightly discounting the future reward, corresponding to a Gamma of 0.9. 

By configuring higher LRs, the agent becomes increasingly biased to take the same action that will enforce the learning policy to be 
particular to a deterministic environment. On the other hand, for very low LRs 
the DQL agent keeps exploring the environment in a complete random behavior without learning. A moderate LR 
provides the equilibrium between a deterministic and stochastic environment. Fig.~\ref{fig:LRComp} compares the learning outcome for three LRs, where a LR of $10^{-4}$ provides the best result. 

\color{black}
Note that one reason of DQN failure is related to the choice of the hyperparameters. 
 There are a number of search techniques that can be used to adjust hyperparameters. In this analysis, we have employed the grid search technique that involves specifying a range of values for each hyperparameter and then training the DQN with all possible combinations of these values. The combination of hyperparameters that produces the best performance is then selected.
\color{black}

\subsection{Learning Performance Evaluation}

\textcolor{black}{Figure~\ref{fig:DQL_QL} 
compares of learning and convergence performance of the proposed DQL scheme with the Q-learning as a benchmark learning algorithm.} 
It plots the total secrecy capacity of the user cluster served by the AR 
over the number of learning episodes for the DQL and Q-learning 
trajectory and power optimization. 
The curves show that the total secrecy capacity of user served by the UAV tends to increase over the episodes until convergence. This validates our approach to define the reward so as to maximize the user rates with unknown CSI of the eavesdropping channel. Initially, 
the total secrecy capacities match for the two algorithms. This is so because of lacking interaction with the environment to provide enough data for training the learning agents. As the learning evolves, favorable actions become more easily discriminated from the unfavorable ones by exploring the environment. 
It is noticeable that the DQL performance substantially exceeds the Q-learning performance due to its ability to approximate the Q-values instead of an inefficient Q-table of the QL, which allows learning gigantic state and action in fewer episodes. 
\begin{figure}[t]
\vspace{+45 pt}
  \subfloat[]{
	\begin{minipage}[c][0.5\width]{
	   0.5\textwidth}
	   \centering
	   \includegraphics[width=0.90\textwidth]{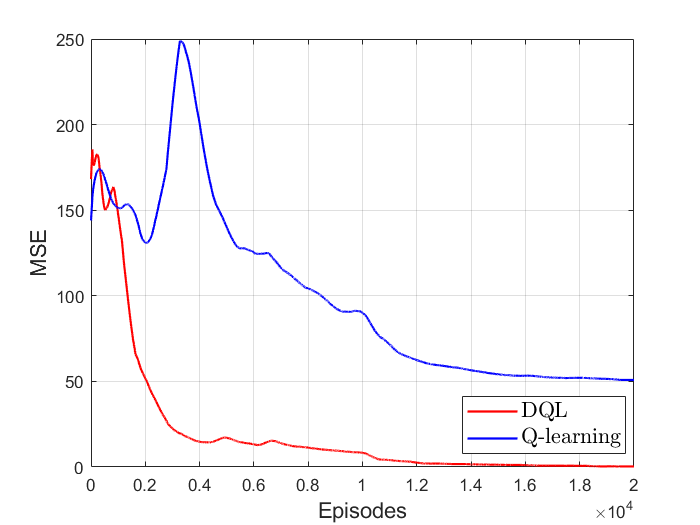}
	    \vspace{+40 pt}
	\end{minipage}}
	\vspace{+41 pt}
 \hfill 
  \subfloat[]
  {
	\begin{minipage}[c][0.5\width]{
	   0.5\textwidth}
	   \centering
	   \includegraphics[width=0.90\textwidth]{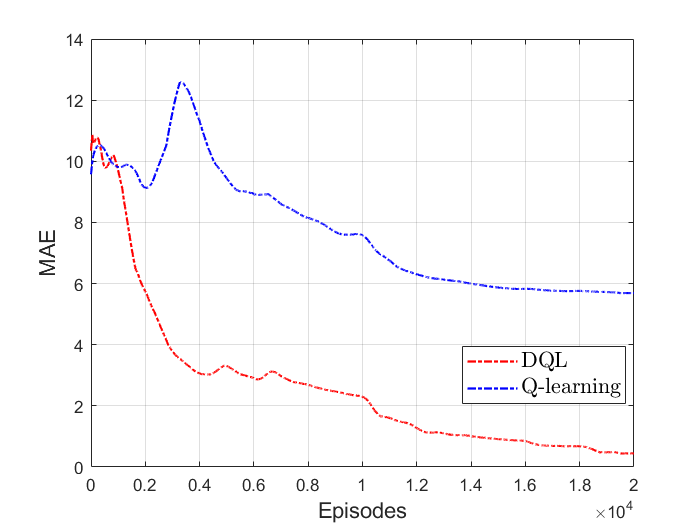}
	   \vspace{+40 pt}
	\end{minipage}}
	\vspace{-5 pt}
\caption{Comparison of the MSE (a) and MAE (b) losses of the DQL and Q-learning \textcolor{black}{over the number of} episodes.
}
\label{fig:MAEloss}
\end{figure}

\textcolor{black}{In order to further analyze the effectiveness and the convergence performance of the learning performance of the DQL design for optimizing the UAV trajectory, 
Fig.~\ref{fig:MAEloss} plots 
the mean square error (MSE) and mean absolute error (MAE) of the DQL and Q-learning solutions over the number of 
episodes. The results show how the loss error is minimized by adjusting the weights of the neural network used to approximate the Q-function. This indicates how well the proposed algorithm is performing and illustrates its convergence toward an optimal policy. 
} The MSE and MAE are calculated by comparing the estimated Q-values using the learned DQL and Q-learning models versus their actually computed values. Those figures reveal 
that the quality of the UAV actions 
are rather poor during the early training phase. As the learning continues, more measurements are accumulated that yield to improved 
actions taken by the UAV agent for both algorithms. The DQL method meets a MSE target of 50 an order of magnitude faster than Q-learning (Fig.~\ref{fig:MAEloss}a). It converges faster and achieves a 35\% higher secrecy capacity after 20000 episodes (Fig.~\ref{fig:DQL_QL}).

\begin{figure}[ht]
	   \centering
	   \includegraphics[width=0.45\textwidth]{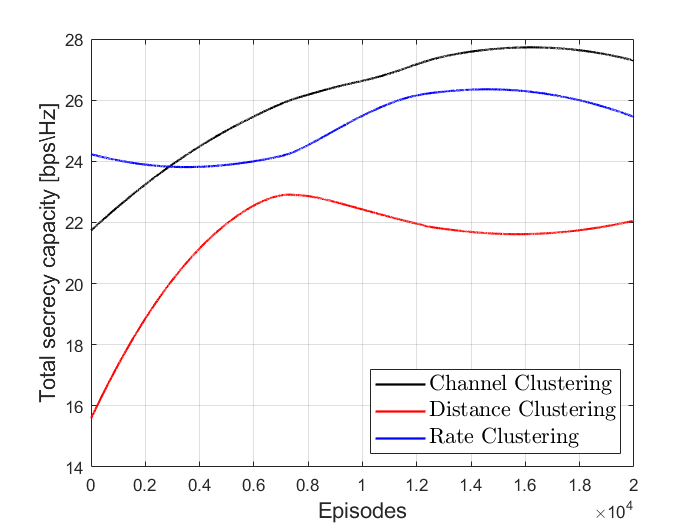}
	    \vspace{-8 pt}
        \caption{Comparison of different clustering techniques. 
        }
        \label{fig:ClusteringComp}
\end{figure}

\subsection{Clustering Performance Evaluation 
}


Here we evaluate the performance of the proposed user clustering scheme and the employed metric on the total secrecy rate performance of the system. 
We consider three metrics for the 
clustering algorithm presented in Algorithm 1: \textit{distance clustering}, where UEs are grouped based on their distances to the GBS, \textit{rate clustering}, where UEs are grouped based on their downlink rates while being served by the GBS, and \textit{channel clustering}, where UEs are grouped based on the normalized channel coefficients. 

Figure~\ref{fig:ClusteringComp}, 
shows the total secrecy capacity of the system after clustering, beamforming, and UAV trajectory optimization for the three clustering metrics. We observe that the proposed channel clustering metric outperforms the rate and distance clustering metrics in terms of total secrecy capacity.

\subsection{Overall Performance Evaluation}
\color{black}
We explain the overall performance evaluation of the proposed method in two parts. In the first part, the impact of the optimal beamforming and power control performance evaluation is studied. In particular, the proposed method is compared with three scenarios where our optimal beamforming and power control are only 
partially implemented. In the second part, the impact of the UAV trajectory on the secrecy capacity is studied. Specifically, the UAV's 3D movement is shown in a scenario in which two clusters of UEs are simultaneously served, one by the UAV, which relocates to best serve the UEs in the cluster, and the other by the GBS.
\color{black}

\textcolor{black}{The context that we study is unique compared to other studies 
as captured in Table I. We develop a framework that involves user clustering, multi-user beamforming, power control, and reinforcement learning for solving the problem and there are no existing studies that propose comparable solution. 
Therefore, we define our own benchmarks to evaluate the proposed framework and the importance of each component comprising it.} The baseline techniques are: 
AR deployment without optimal GBS beamforming (UAV+NoBF), no AR deployment with optimal GBS beamforming (NoUAV+BF), and 
no AR deployment and without optimal GBS beamforming (NoUAV+NoBF). \textcolor{black}{In all cases where the AR is deployed the optimal beamforming and power control of the AR is activated.} 

\begin{figure}[t]
	   \centering
	   \includegraphics[width=0.45\textwidth]{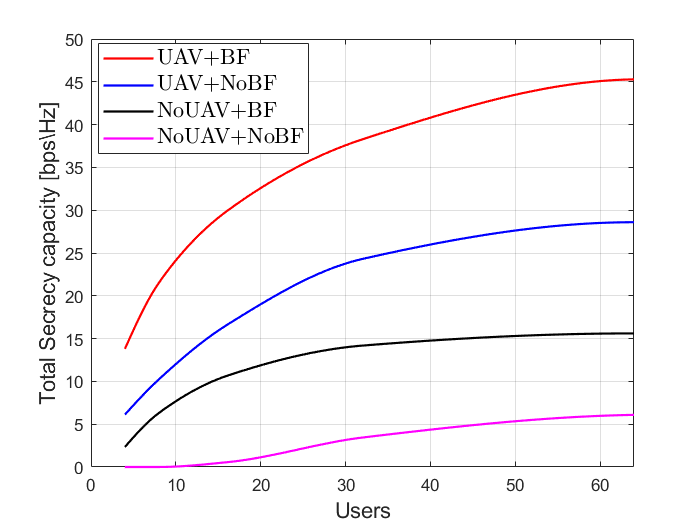}
	    \vspace{-8 pt}
        \caption{Comparison of different techniques Vs number of users for total secrecy capacity of the system.
        }
        \label{fig:TechComp}
\end{figure}

Figure~\ref{fig:TechComp} shows the achieved total secrecy capacity 
over the 
number of users. 
The secrecy capacity 
improves with the number of users for all schemes. 
The 
proposed solution clearly outperforms the other techniques. 
The UAV+NoBF scheme achieves a better secrecy capacity 
than the NoUAV+BF scheme. That is, deploying an AR is more useful for improving the secrecy capacity 
than employing optimal multi-user beamforming at the GBS. 
Nevertheless, beamforming and power control schemes have a notable contribution to the secrecy performance of the system as can be observed when comparing the performance of the NoUAV+NoBF scheme with the proposed solution and other benchmark techniques.  The optimal beamforming and power control 
increases the SNR and reduces the multi-user interference 
while minimizing the likelihood of eavesdropping and improving the overall secrecy capacity. 
The addition of the UAV as an AR allows to serve those users effectively that have a worse channel to the GBS. This is accomplished by the proposed clustering method and the UAV trajectory optimization along with beamforming and power control. 

\begin{figure*}[h!]
\vspace{-12 pt}
	   \centering
	   \includegraphics[width=0.85\textwidth]
    {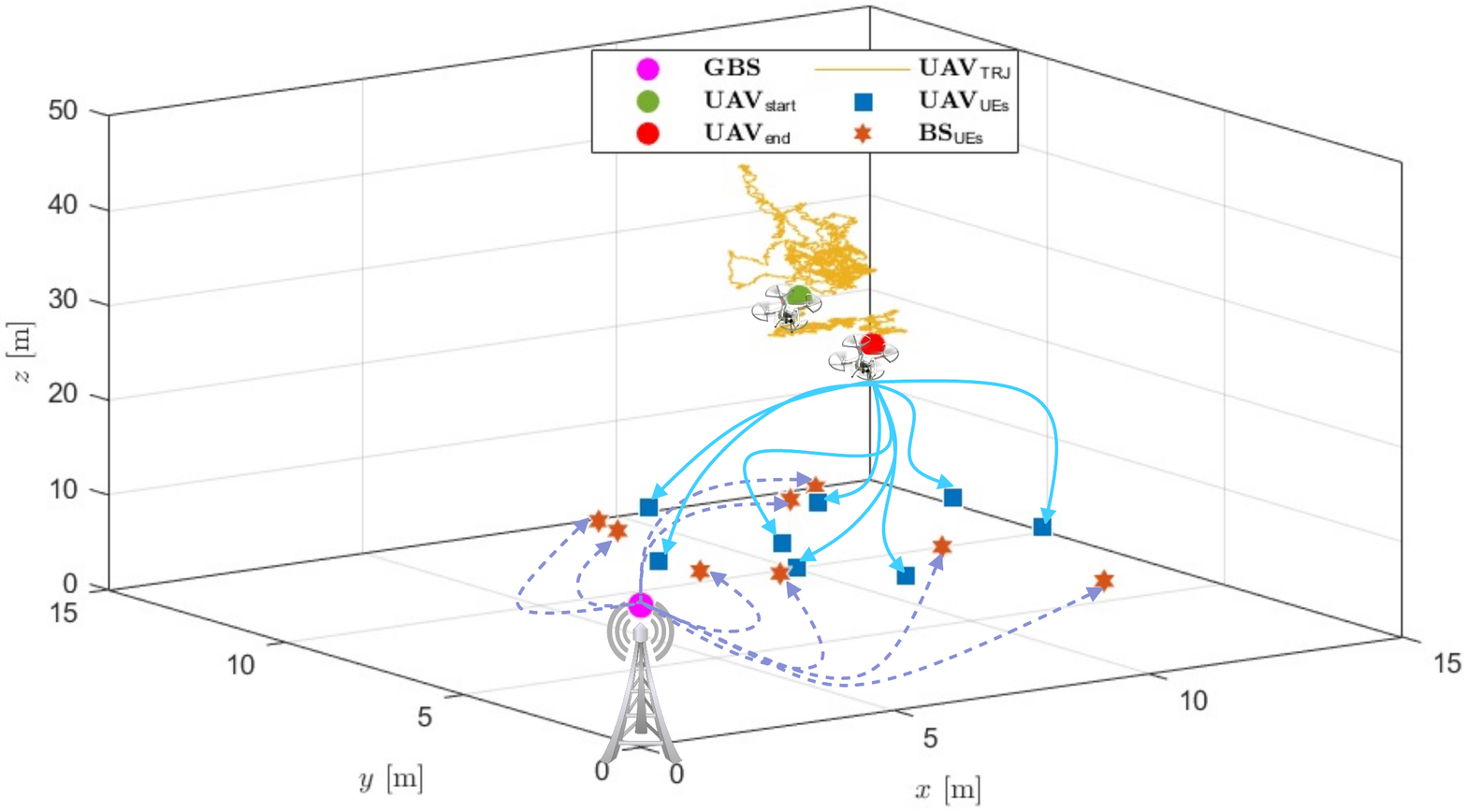}
	    \vspace{-3mm}
        \caption{Illustration of channel based user clustering of Algorithm 1 and UAV trajectory  optimization of Algorithm 2. 
        }
        \vspace{-6 mm}
        \label{fig:Trajectory}
\end{figure*}
\begin{figure}[ht]
	   \centering
	   \includegraphics[width=0.45\textwidth]{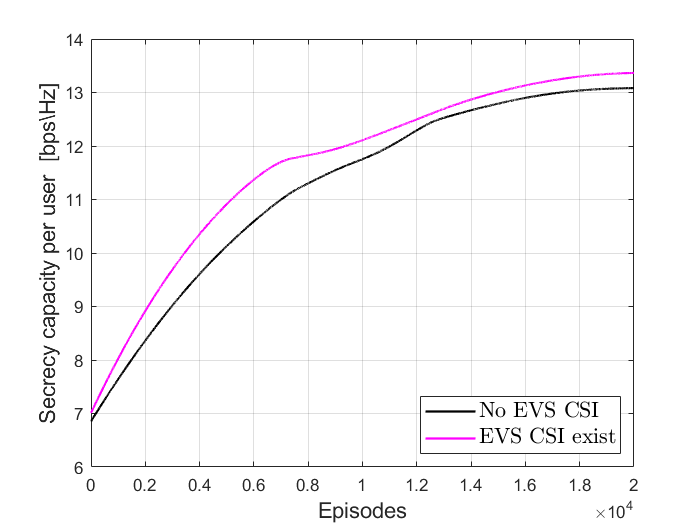}
\vspace{-3mm}
\caption{\color{black}Secrecy capacity 
of the proposed optimization framework 
for the cases of known and unknown CSI of eavesdroppers, employing the secrecy capacity 
and user capacity as the reward function, respectively.}
\label{fig:Comparison}
\end{figure}

Figure~\ref{fig:Trajectory} illustrates the dynamic 3D trajectory optimization process 
for the case of 16 users where there are 8 users in each cluster as a result of Algorithm 1 that employs channel based clustering. 
\textcolor{black}{We observe that the UAV moves toward the center of the area where the users are located, and its final position 
is near the minimum height to be as close as possible to the UEs served by the AR for ensuring good channels to be able to lower the transmission power and thus increase the secrecy capacity, 
while also ensuring that the UAV stays within its operational limits and avoids ground obstacles. Overall, the dynamic 3D trajectory optimization process, combined with 
optimal multi-user beamforming and power control, helps achieve high secrecy capacities 
(Fig. 9) in the presence of eavesdroppers.}

\color{black}
\subsection{Known vs. Unknown Information of Eavesdroppers}
In order to put our contribution in context and provide further justification for our optimization framework, 
we consider the case where the location and the channel states of eavesdroppers are known to the GBS and the UAV in the proposed communication context. 
Knowning the CSI of eavesdropping channels allows employing the secrecy capacity 
(24) as the reward function. 

We simulate 16 users 
that are clustered in two groups to be served by the GBS or AR resulting from the channel based clustering with known eavesdropper locations and CSI. 
Figure 11 shows the resulting average secrecy capacity 
per user with and without eavesdropping information available to the network. For the case of unknown eavesdropping channels, we employ the proposed optimization solution and reward function based on the legitimate user rate. 
The UAV adjusts its power and trajectory according to the available information. 
As expected, having the 
information of malicious actors eavesdropping on the wireless links allows the network to adjust its parameters better and increase the secrecy capacity. 

Figure 11 also indicates that the secrecy capacity 
performance gap of not knowing the channel characteristics 
of eavesdroppers is not significant. In other words, blindly optimizing the secrecy capacity 
by focusing on the legitimate user rates produces an outcome that is very close to an optimization framework that has and leverages the full information about eavesdroppers. 
The reason for this 
is 
that the proposed practical solution with unknown CSI imperceptibly considers the possible CSI between the base stations and the eavesdroppers. 

We conclude that 
despite the practical assumption of not knowing the CSI of eavesdropping channels and not exploring methods other than optimizing the user rates, the proposition of this paper, the proposed communications and optimization framework can accomplish a performance that is very close to a network that has access to the full information about eavesdroppers. This encourages doing further research on 
improving the proposed technique, for example, by considering partially known information about eavesdroppers or other reasonable assumptions, or even exploring new physical layer security metrics.

\subsection{User Mobility}
In this subsection, 
we examine the secrecy capacity 
of mobile users with a static eavesdropper.
Without loss of generality, the mobility of the ground users will be over the x-axis with a fixed y-position. We define the distance step parameter ($dx$), 
which corresponds to the granularity of movement. Then, $UE_{X_C^{t+1}} = UE_{X_C^{t}} + dx$, 
can be defined to model the mobility of the ground users, 
\begin{figure}[ht]
	   \centering
	   \includegraphics[width=0.45\textwidth]{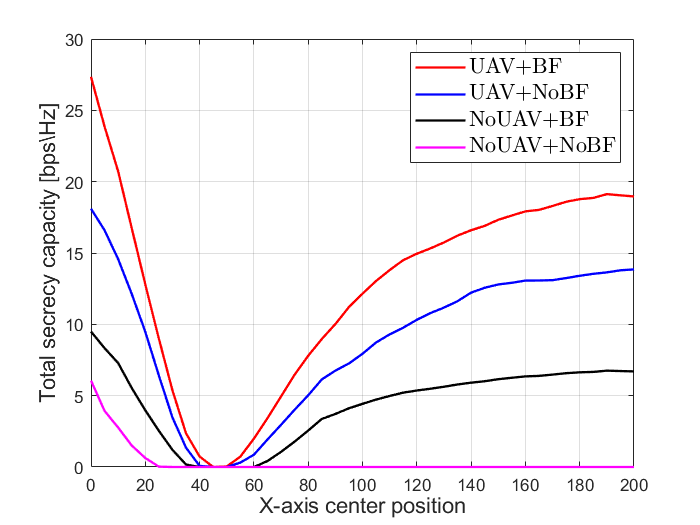}
\vspace{-3mm}
\caption{\color{black}Comparing the performance of the proposed solution under user mobility scenario with other benchmark techniques.}
\label{fig:MOB}
\end{figure}
where $UE_{X_C^{t+1}}$ is the next center x-positions of all the ground users in the next time step and for each new center the users are redistributed randomly around the center. This process enables the users to simulate a realistic movement pattern that reflects their actual movement. 

We consider 16 ground users to be served either directly by the GBS or 
through the AR. Additionally, with each movement of step $dx$ , the proposed solution of Fig.~2 
re-clusters the users, re-performs beamforming and power control for the GBS and UAV transmissions, and re-optimizes the trajectory of the UAV given the new positions of users. Figure~\ref{fig:MOB} presents the obtained total secrecy capacity 
over the 
center position of the moving user cluster 
for the proposed solution and the benchmarks introduced in Fig. \ref{fig:TechComp}. 

The results of Fig.~\ref{fig:MOB} show 
that the proposed solution achieves a higher total secrecy capacity 
compared to the other schemes. By optimizing the trajectory of the UAV, the system ensures that it flies as close as possible to the users being served by the AR enabling good channels and lower transmit powers. 
We observe that the secrecy capacity of the proposed solution drops to zero only for the case where 
the center of the user cluster matches the eavesdropper position. After the users pass the eavesdropper, the secrecy capacity 
rapidly recovers. 

Notice the secrecy capacity 
after passing the eavesdroppers is lower than before reaching it. This is because of the lower data rates achieved by the direct GBS links experiencing a higher path loss with increasing distance. On the other hand, the NoUAV+NoBF scheme has reached the zero secrecy capacity 
much earlier and remains at this state even after leaving the eavesdropper behind. When comparing the performance of the UAV+NoBF and the NoUAV+BF schemes, we again realize the effectiveness of deploying the AR for achieving a higher secrecy capacity. 

\color{black}

\section{Conclusions}
\vspace{-1 mm}
\label{sec:conclusions}


This paper addressed the major eavesdropping problem in present-day wireless communications. We developed a practical framework against passive eavesdroppers in multi-user cellular networks without knowledge of the eavesdroppers’ locations and CSI channels. Considering the unknowns, we optimized the user rates employing advanced wireless techniques at the physical layer to improve the sum-secrecy capacity among all users in a cell. Specifically, we suggested employing multi-user beamforming and deploying a UAV that serves as an AR. We clustered the users into two groups wherein users are either served by the GBS or by the AR, whose 3D position, multiuser beamforming matrix, and transmit powers are optimized combining closed-form expressions with machine learning techniques. Specifically, we designed and analyzed a DQN for the UAV trajectory optimization subproblem. Numerical results showed that the proposed system achieves highest secrecy capacities and scales well over the number of users to be served.

\color{black}
Lessons learned from this work can lead to a number of research directions for solving open research challenges. 
\textcolor{black}{
We will examine additional UAV specific operational constraints, including energy consumption, flight time,
and speed, in future work.} 
\textcolor{black}{These are especially important for implementation and deployments of ARs with today's small UAVs. One can prototype and validate the presented techniques on the Aerial Experimentation and Research Platform for Advanced Wireless (AERPAW)~\cite{AerpawAly}, which facilitates implementing the proposed communications system 
with software radios 
and conducting different types of mobility experiments by leveraging AERPAW’s unmanned ground vehicles. 
}

\color{black}

\section*{\color{black} Appendix}
\vspace{-1 mm}
\label{sec:Appendix}
\color{black}

\subsection{Proof of Lemma \ref{lem:objective}}
\begin{proof}
\color{black}
The substitution of $\vb*{H}_1$, $\vb*{H}_2$, $\vb*{W}_{br}$, and $\vb*{W}_r$, which are defined in (\ref{eq:pwr_cntrl_2}), (\ref{eq:pwr_cntrl_3}), (\ref{eq:pwr_cntrl_4}), and (\ref{eq:pwr_cntrl_5}), respectively, in $\vb*{y}_2$ defined in (\ref{eq:pwr_cntrl_1}) yields
\begin{align}
\notag    \vb*{y}_2 &=  \vb*{U}_2 \, \vb*{\Sigma}_2 \, \underbrace{ \vb*{V}_2 \ssymbol{2} \, \vb*{V}_2 }_{\vb*{I}}\, \hat{\vb*{\Lambda}}_r \, \vb*{U}_1 \ssymbol{2} \,\vb*{\Lambda}_r \,  \vb*{U}_1 \,\vb*{\Sigma}_1 \,\underbrace{ \vb*{V}_1 \ssymbol{2} \, \vb*{V}_1}_{\vb*{I}}\, \vb*{\Lambda}_b \, \vb*{U}_2 \ssymbol{2} \, \vb*{s}_K \, \\
 & \ + \, \vb*{U}_2 \, \vb*{\Sigma}_2 \,   \underbrace{ \vb*{V}_2 \ssymbol{2} \, \vb*{V}_2 }_{\vb*{I}}\,         \hat{\vb*{\Lambda}}_r \, \vb*{U}_1 \ssymbol{2} \, \vb*{\Lambda}_r \, \vb*{n}_1 \, + \, \vb*{n}_2. 
\end{align}
Therefore we have
\begin{align} \label{eq:pwr_cntrl_6}
\notag    \vb*{y}_2 &=  \underbrace{\vb*{U}_2 \, \vb*{\Sigma}_2 \, \hat{\vb*{\Lambda}}_r \, \vb*{U}_1 \ssymbol{2} \,}_{\vb*{I}}  \vb*{\Lambda}_r \,  \underbrace{\vb*{U}_1 \, \vb*{\Sigma}_1 \, \vb*{\Lambda}_b \, \vb*{U}_2 \ssymbol{2}}_{\vb*{I}}   \, \vb*{s}_K \, \\[-4pt]
    & \quad + \, \vb*{U}_2 \, \vb*{\Sigma}_2 \, \hat{\vb*{\Lambda}}_r \, \vb*{U}_1 \ssymbol{2} \, \vb*{\Lambda}_r \, \vb*{n}_1 \, + \, \vb*{n}_2,\\[-22pt]
               \notag
\end{align}
where the matrices $\vb*{I}$ are obtained using the ZF criterion, i.e., $\vb*{U}_2 \, \vb*{\Sigma}_2 \, \hat{\vb*{\Lambda}}_r \, \vb*{U}_1 \ssymbol{2} = \vb*{I}$ and $\vb*{U}_1 \, \vb*{\Sigma}_1 \, \vb*{\Lambda}_b \, \vb*{U}_2 \ssymbol{2} = \vb*{I}$. As a result, the simplified equation is  
\color{black}
\vspace{-8pt}
\begin{align}\label{eq:pwr_cntrl_9}
        \vb*{y}_2 &= \vb*{\Lambda}_r \, \vb*{s}_K \, + \, \vb*{\Lambda}_r \, \vb*{n}_1 \, + \, \vb*{n}_2\\[-22pt]
               \notag
\end{align}
\color{black}
in which
\begin{equation*}
\vb*{\Lambda}_{r} = 
\begin{pmatrix}
    \lambda_{r,1} & 0 & \ldots & 0 & 0 \\
    \vdots & \vdots & \ldots &  \vdots & 0 \\
    0 & \ldots &  \lambda_{r,k} & 0 & 0 \\
    0 & \vdots & \ldots &  \vdots & 0 \\
    0 & 0 & \ldots & 0 & \lambda_{r,K}
\end{pmatrix}_{K\times K}
,
\vb*{s}_{K} =
\begin{pmatrix}
    s_{1,1} \\
    \vdots \\
    s_{k,1} \\
    \vdots \\
    s_{K,1}
\end{pmatrix}_{K\times 1}.
\end{equation*}
$\vb*{\Lambda}_{r}$ is a diagonal matrix and $\vb*{n}_1, \vb*{n}_2 \in \mathbb{C}^{K\times 1}$ are the noise vectors. 
The simplified SINR can then be written 
as
\begin{align}\label{eq:pwr_cntrl_1001}
         SINR = \frac{\lambda_{r,k}^2}{\lambda_{r,k}^2+1},
\end{align}
which proves Lemma IV.1.
\color{black}

\color{black}

\end{proof}
\subsection{Proof of Lemma \ref{lem:const}}
\begin{proof}
The beamforming power at the relay can be simplified as follows
\vspace{-8pt}
\color{black}
\begin{align}
             P_r &= Tr \big( \vb*{x}_r \vb*{x}_r^{\dagger} \big) \label{eq:apdx_0} \\[-2pt]
                 &= Tr \bigg( \vb*{W}_r \Big( \underbrace{\vb*{H}_1 \vb*{W}_b \vb*{W}_b^{\dagger} \vb*{H}_1^{\dagger}}_{term\ i} + \vb*{I}   \Big) \vb*{W}_r^{\dagger} \bigg),\label{eq:apdx_1} \\[-7pt] 
                 &= Tr \bigg( \vb*{W}_r \Big( \underbrace{\vb*{U}_1 \vb*{\Sigma}_1 \vb*{\Lambda}_b \vb*{\Lambda}_b^{\dagger} \vb*{\Sigma}_1^{\dagger}  \vb*{U}_1^{\dagger}}_{\vb*{I}} + \vb*{I}   \Big) \vb*{W}_r^{\dagger} \bigg), \label{eq:apdx_2}\\[-5pt] 
                 &= 2 \times Tr \Big( \vb*{W}_r \, \vb*{W}_r^{\dagger} \Big)\label{eq:apdx_four}\\[-22pt]
               \notag
\end{align}

\color{black}
where (\ref{eq:apdx_1}) is obtained by replacing (\ref{eq:relay_1}) and (\ref{eq:relay_2}) in (\ref{eq:apdx_0}), (\ref{eq:apdx_2}) is derived by substituting (\ref{eq:pwr_cntrl_2}) and (\ref{eq:pwr_cntrl_4}) in \textit{term i} of (\ref{eq:apdx_1}), \color{black} and (54) is obtained from $\vb*{U}_1 \, \vb*{\Sigma}_1 \, \vb*{\Lambda}_b \, \vb*{U}_2 \ssymbol{2} = \vb*{I}$. \color{black}
Subsequently, by replacing (\ref{eq:pwr_cntrl_5}) into (\ref{eq:apdx_four}), we have
\vspace{-6pt}
\begin{align}
\notag  P_r &= 2 \times Tr \Big( \vb*{W}_r \, \vb*{W}_r^{\dagger} \Big) \\[-6pt]
            &= 2 \times Tr \Big(  \vb*{V}_2 \, \underbrace{\hat{\vb*{\Lambda}}_r \, \vb*{U}_1 \ssymbol{2}}_{term\ ii}  \, \vb*{\Lambda}_r^2  \, \underbrace{\vb*{U}_1 \, \hat{\vb*{\Lambda}}_r}_{term\ iii}  \, \vb*{V}_2 \ssymbol{2}   \Big) \label{eq:apdx_5} \\[-4pt]
            &= 2 \times Tr \Big(  \vb*{V}_2  \vb*{\Sigma}_2^{-1}  \vb*{U}_2 \ssymbol{2}
            \vb*{\Lambda}_r^2 \vb*{U}_2 \vb*{\Sigma}_2^{-1} \vb*{V}_2 \ssymbol{2} \Big) \label{eq:apdx_6} \\[-4pt]
            &=  2 \, \sum\limits_{m=1}^K  \sum\limits_{n=1}^K |\vb*{U}_2(m,n)|^2 \, \sigma_{2,n}^{-2} \, \lambda_{r,m}^2,\\[-22pt]
               \notag
\end{align}
\color{black} where considering $\vb*{U}_2 \, \vb*{\Sigma}_2 \, \hat{\vb*{\Lambda}}_r \, \vb*{U}_1 \ssymbol{2} = \vb*{I}$ for\textit{term ii} and \textit{term iii} of (\ref{eq:apdx_5}) yields (\ref{eq:apdx_6}). \color{black}
\end{proof}

\subsection{Proof of Theorem \ref{thm:optimization_1}}
\begin{proof}
We need to obtain the optimum beamforming power elements and Lagrange multipliers, i.e., $\lambda_{r,l}^{*}$, $\alpha_1^{*}$, $\alpha_{2,l}^{*}$, and $\alpha_{3,l}^{*}$, where $l = \{1, \cdots, K\}$.
To this end, we apply the Karush-Kuhn-Tucker (KKT) conditions to this problem as has been applied to similar ones \cite{Boyd} \cite{Guizani} \cite{ICC21wkshp}. From the gradient condition and the complementary slackness condition, we have
\vspace{-8pt}
\begin{align}
& \nabla_{\lambda_{r,l}} \mathcal{L}(\lambda_{r,l}^{*}, \alpha_1^{*}, \alpha_{2,l}^{*}, \alpha_{3,l}^{*}) = 0, \label{eq:pwr_cntrl_14}\\[-3pt]
& - \alpha_1^{*} \bigg( 2 \, \sum\limits_{l=1}^K  \sum\limits_{n=1}^K |\vb*{U}_2(l,n)|^2 \, \sigma_{2,n}^{-2} \, \lambda_{r,l}^{*\ 2} - P_{r, max}  \bigg) = 0, \label{eq:pwr_cntrl_15}\\[-4pt]
& - \alpha_{2,l}^{*} \Big( \lambda_{r,l}^{*} - \lambda_{r, max} \Big) =0, \label{eq:pwr_cntrl_16}\\[-4pt]
& - \alpha_{3,l}^{*} \Big( - \lambda_{r, l}^{*} \Big) =0.\label{eq:pwr_cntrl_17}\\[-22pt]
               \notag
\end{align}

By simplifying (\ref{eq:pwr_cntrl_14}), we obtain
\vspace{-4pt}
\begin{align}\label{eq:pwr_cntrl_18}
\notag & \frac{2\, \lambda_{r,l}^{*}}{\big(2\lambda_{r,l}^{* \, 2} + 1\big)\, \big(\lambda_{r,l}^{* \, 2} + 1\big)\, \text{ln}\, 2} \ - 4 \, \alpha_1^{*} \, \lambda_{r,l}^{*}  \\[-2pt]
& \times \, \sum\limits_{n=1}^K |\vb*{U}_2(l,n)|^2 \, \sigma_{2,n}^{-2} - \, \alpha_{2,l}^{*} \, + \, \alpha_{3,l}^{*} = 0.\\[-22pt]
               \notag
\end{align}
Applying the KKT conditions yields the optimal beamforming power as follow
\vspace{-8pt}
\begin{equation*}
\lambda_{r,l}^{*}=
\begin{cases}
\small          0 \qquad    &\quad  \lambda_{r,l}^{\dagger} \leq 0, (\alpha_1^{*} \, F\,  \text{ln}\,2)>0.25 \\
          \lambda_{r,l}^{\dagger} \qquad   &\qquad 0 < \lambda_{r,l}^{\dagger} < \lambda_{r,max}\\
          \lambda_{r,max} \qquad  &\qquad \quad \ \ \lambda_{r,l}^{\dagger} \geq \lambda_{r,max}
\end{cases}
\vspace{-7pt}
\end{equation*}
in which
\begin{align} \label{eq:appendx_lambda}
\vspace{-5pt}
   \lambda_{r,l}^{\dagger} = \sqrt{\frac{1}{4}\Big(\sqrt{1+\frac{2}{\alpha_1^{*} \, F \, \text{ln}\,2 }} - 3\Big)},\\[-22pt]
               \notag
\end{align}
where $F$ is a constant that equals to $\sum\limits_{n=1}^K |\vb*{U}_2(l,n)|^2 \, \sigma_{2,n}^{-2}$, and the Lagrangian multiplier $\alpha_1^{*}$ can be obtained by replacing (\ref{eq:pwr_cntrl_19}) into the first constraint of (\ref{eq:pwr_cntrl_12_1}) when the equality holds: \color{black} $\alpha_1^{*} = f_1(P_{r, max}, \vb*{U}_2, \Sigma_{2})$. \color{black} 
Note that if the last constraint defined in the condition (\ref{eq:pwr_cntrl_17}) is binding, i.e., if $\lambda_{r, l}^{*}=0$, then $\alpha_{1}^{*}= \alpha_{2,l}^{*}= 0$ due to the complementary slackness conditions. Substituting these multipliers in (\ref{eq:pwr_cntrl_14}) results in $\alpha_{3,l}^{*}=0$. Also, replacing $\lambda_{r, l}^{*}=0$ in the objective function of (\ref{eq:pwr_cntrl_12}) results in a zero capacity rate, which is not desired. In the same way as in \cite{Guizani, ICC21wkshp}, it can be considered that $\lambda_{r, l}^{*} = \lambda_{r, max}$ for the values of beamforming powers above than the maximum. 
\color{black}
In addition, the value of $\lambda_{r,l}^{\dagger}$ in (\ref{eq:appendx_lambda}) can be numerically obtained for different values of channel coefficients and the UAV's power limitation, i.e., $\vb*{U}_2$, $\sigma_{2,n}^{-2}$, and $P_{r, max}$. However, using the Taylor series in (\ref{eq:appendx_lambda}) at $x =\frac{2}{\alpha_1^{*} \, F \, \text{ln}\,2 }$ with negligible $O(x^2)$, one can further simplify the obtained $\lambda_{r,l}^{\dagger}$ as
\begin{align} 
\vspace{-5pt}
   \lambda_{r,l}^{\dagger 2} = \frac{1}{4 \, f_1 \, \sum\limits_{n=1}^K |\vb*{U}_2(l,n)|^2 \, \sigma_{2,n}^{-2} \ \text{ln}\,2 } - 0.5,\\[-22pt]
               \notag
\end{align}
where $\lambda_{r,l}^{\dagger}$ is a positive number and $f_1$ is the above defined function of the UAV power and channel coefficients.
\color{black}
Finally, it is worth pointing out that if the UAV uses only one antenna for communicating with its users, then the beamforming power for the antenna is set to be $\lambda_{r,max}$.
\end{proof}

\color{black}


\balance

\begin{IEEEbiography}[{\includegraphics[width=1in,height=1.25in,clip,keepaspectratio]{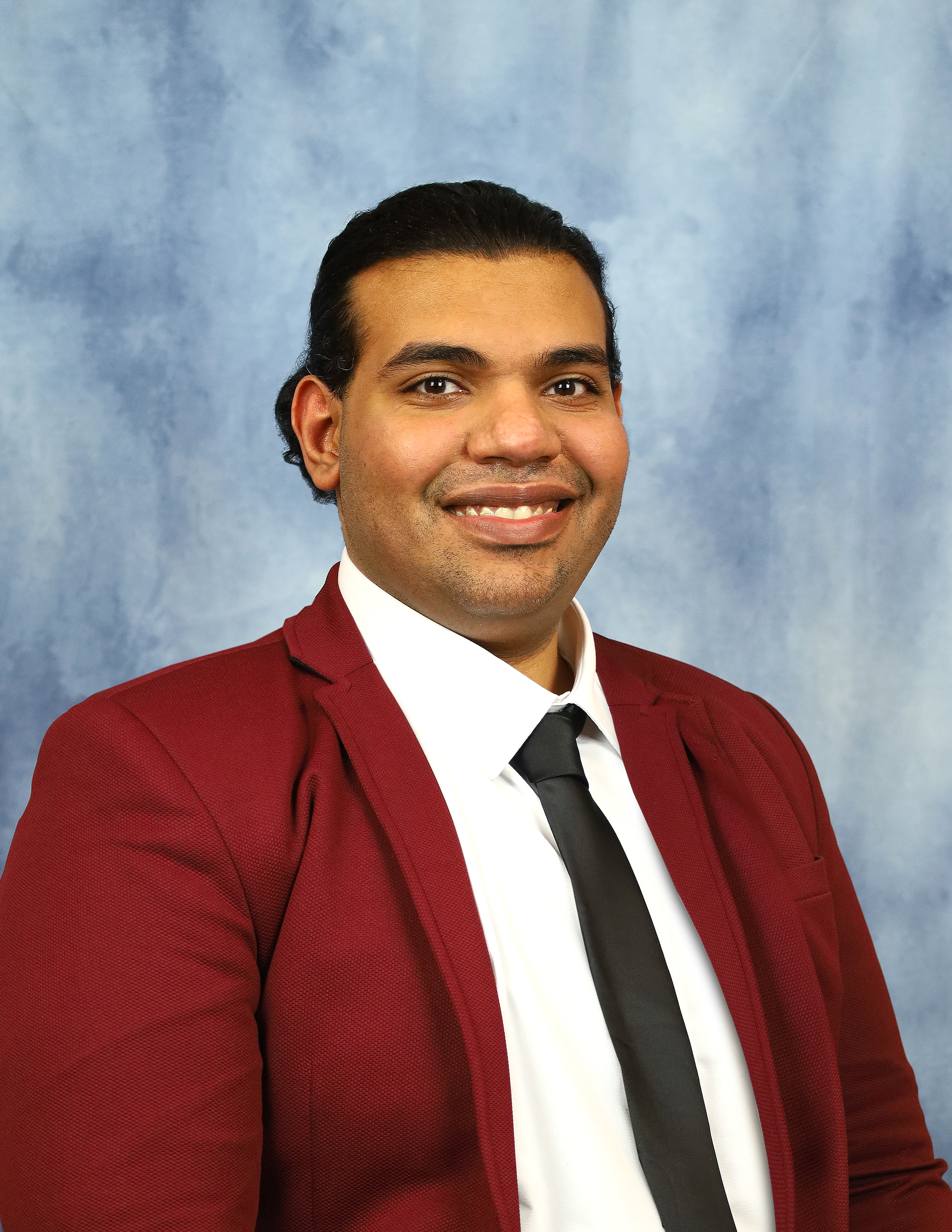}}]{Aly Sabri Abdalla} received the Ph.D. degree in  Electrical and Computer Engineering at Mississippi State University, MS, USA in 2023. He received the B.S. and M.S. degrees in Electronics and Communications Engineering from the Arab Academy for Science Technology and Maritime Transport, Egypt, in 2014 and 2019, respectively. 

Since 2019 
he has been a Research Assistant in the Department of Electrical and Computer Engineering at Mississippi State University. His research interests include 
wireless communication and networking, software radio, spectrum sharing, wireless testbeds and testing, and wireless security with application to mission-critical communications, open radio access network (O-RAN), unmanned aerial vehicles (UAVs), and reconfigurable intelligent surfaces (RISs). He is serving as a student member of the IEEE Vehicular Technology Society’s Ad Hoc Committee on Drones and IEEE 1920.1: Aerial Communications and Networking.
\end{IEEEbiography}

\begin{IEEEbiography}[{\includegraphics[width=1in,height=1.25in,clip,keepaspectratio]{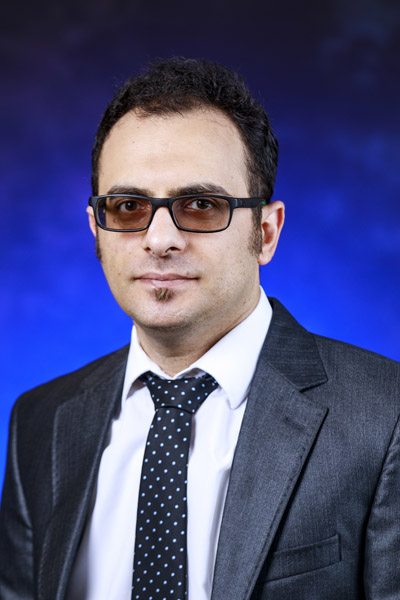}}]{Ali Behfarnia} received the Ph.D. degree in electrical engineering and computer science from Wichita State University, KS, USA in 2020. He is currently serving as an Assistant Professor in the Department of Engineering with the University of Tennessee at Martin. His research interests include wireless communication systems, game theory, error correction coding, cyber-physical systems, and the applications of machine and deep learning in communications over wireless networks.
\end{IEEEbiography}

\begin{IEEEbiography}[{\includegraphics[width=1in,height=1.25in,clip,keepaspectratio]{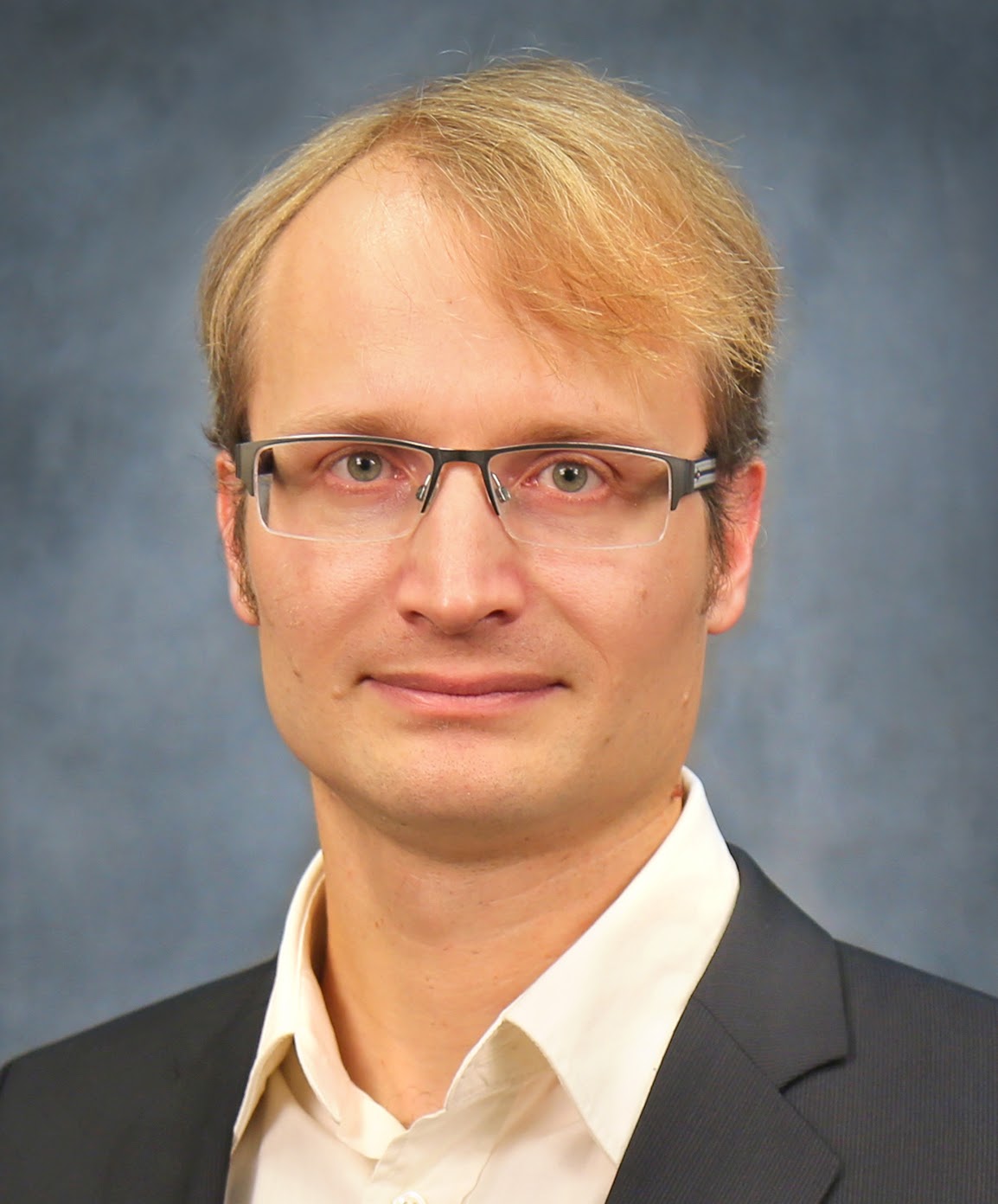}}]{{Vuk Marojevic}} (Senior Member, IEEE)
received the M.S. degree from the University of
Hanover, Germany, in 2003, and the Ph.D. degree
from Barcelona Tech-UPC, Spain, in 2009, all in
electrical engineering. 

He is currently an Associate
Professor with the Department of Electrical and
Computer Engineering, Mississippi State University. His research interests include 4G/5G security, spectrum sharing, software radios, testbeds,
resource management, and vehicular and aerial
communications technologies and systems. 

Prof. Marojevic is an Editor of the IEEE
Transactions on Vehicular Technology, an Associate Editor of IEEE
Vehicular Technology Magazine, and an Officer of the IEEE ComSoc Aerial
Communications Emerging Technology Initiative.
\end{IEEEbiography}

\vspace{0.2cm}
\noindent

\vspace{0.2cm}
\noindent

\vspace{0.2cm}
\noindent

\end{document}